\documentclass[12pt]{iopart}

\usepackage{graphicx}
\usepackage{iopams}
\begin{document}

\title[Phase diagram of the fully anisotropic two-leg $XXZ$ spin ladder]
{Groundstate fidelity phase diagram of the fully anisotropic two-leg spin-$\case12$ $XXZ$ ladder}

\author{Sheng-Hao Li$^{1,2}$, Qian-Qian Shi$^{1,3}$, Murray T. Batchelor$^{1,4}$ and Huan-Qiang Zhou$^1$}
\address{$^1$ Centre for Modern Physics and Department of
Physics, Chongqing University, Chongqing 400044, The People's
Republic of China}
 \address{$^2$ Chongqing Vocational Institute of Engineering, Chongqing 402260, The People's Republic of China}

\address{$^3$ College of Materials Science and Engineering and Centre for Modern Physics,
Chongqing University, Chongqing 400044, The People's
Republic of China}

\address{$^4$ Mathematical Sciences Institute and Department of
 Theoretical Physics, Research School of Physics and Engineering,
 Australian National University, Canberra ACT 2601, Australia}


\begin{abstract}
The fully anisotropic two-leg spin-$\case12$ $XXZ$  ladder model is studied in terms of an algorithm
based on the tensor network representation of quantum many-body states
as an adaptation of projected entangled pair states to the geometry of translationally invariant infinite-size quantum spin ladders.
The  tensor network algorithm provides an effective method to generate the groundstate wave function, which
allows computation of the groundstate fidelity per lattice site, a universal marker to detect phase transitions in quantum many-body systems.
The groundstate fidelity is used in conjunction with local order and string order parameters to systematically
map out the groundstate phase diagram of the ladder model.
The phase diagram exhibits a rich diversity of quantum phases.
These are the ferromagnetic, stripe ferromagnetic,  rung singlet, rung triplet,  N\'eel,
stripe N\'eel and Haldane phases, along with the two $XY$ phases $XY1$ and $XY2$.
\end{abstract}

%
\vspace{2pc}

\section{Introduction}

Spin ladder systems have attracted considerable attention from both experimentalists and theoreticians alike~\cite{ED, review}.
One of the most striking features of the simple spin-$\case12$ Heisenberg ladder is that the spin excitations are gapful (gapless) when the
number of legs is even (odd)~\cite{White}.
Spin ladder systems in general represent a particularly interesting class of
quantum critical phenomena, exhibiting a rich variety of quantum phases~\cite{phase1,phase2,phase3,phase4,phase5}.
Apart from a few cases~\cite{review}, spin ladder systems are not exactly solvable,
therefore it is necessary to develop various techniques, both analytical and numerical,
to investigate their physical properties~\cite{p1,p2,p3,p4,p5,p6,p7,p8,p9,p10,p12}.

An efficient tensor network (TN) algorithm has been developed which is tailored to
translationally invariant infinite-size spin ladder systems~\cite{lsh}.
The algorithm is based on an adaptation to the ladder geometry of projected entangled pair states~\cite{peps},
a representation of quantum many-body wave functions.
The spin ladder TN algorithm has been successfully applied to a variety of models~\cite{lsh,cross}, including the ferromagnetic
frustrated two-leg ladder, the two-leg Heisenberg spin ladder with cyclic four-spin exchange and cross couplings, and the
three-leg Heisenberg spin ladder with staggering dimerization.
The algorithm is seen to be efficient compared to the density matrix renormalization group~\cite{p4}
and time evolving block decimation~\cite{vidal}, at least as far as the memory cost is concerned.

The TN representation offers an efficient way to detect quantum phase transitions in many-body quantum systems~\cite{QPT1,QPT2}
using the groundstate fidelity per lattice site~\cite{b6,zhou,b7,zhou1,b3,whl}
(for other work on the fidelity approach to quantum phase transitions, see also~\cite{b0,fidelity1,you,rams}).
We recall that fidelity, as a measure of quantum state distinguishability in
quantum information science, describes the distance between two given quantum states.
It offers a powerful means to investigate critical phenomena in quantum many-body lattice systems,
as demonstrated in Refs.~\cite{b6,zhou,b7,zhou1,b3,whl}.
For example, the groundstate fidelity per lattice site enables the characterization of drastic changes of the
groundstate wave functions in the vicinity of a phase transition point.
A systematic scheme to study critical phenomena in quantum many-body lattice systems has been outlined using this approach~\cite{zhou-op}.
Once the groundstate phase diagram is mapped out by means of the groundstate fidelity per lattice site,
the local order parameters can be derived from the reduced density matrices for a representative groundstate
wave function in a symmetry-broken phase.
Other phases without any long-range order can also be detected and further characterized by
nontrival order parameters such as string order and pseudo-order parameters.

\begin{figure}
\begin{center}
\includegraphics[width=0.55\textwidth]{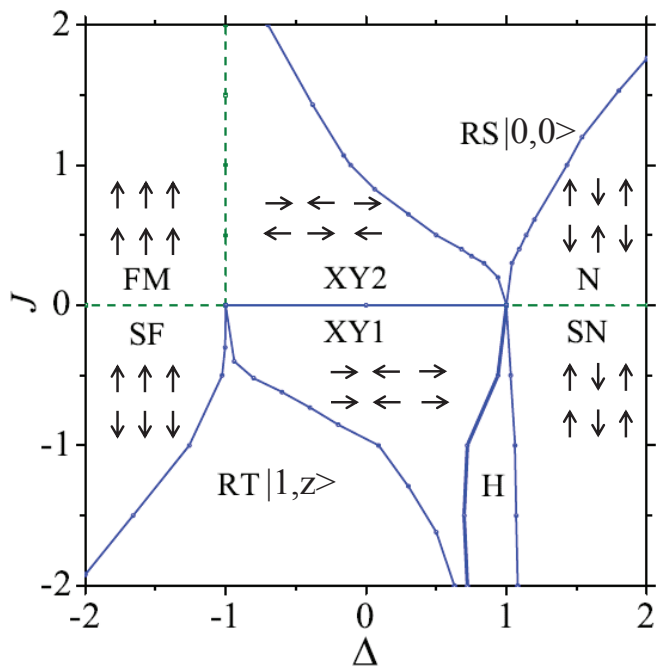}
\end{center}
\caption{The groundstate phase diagram of the spin-$\case12$ $XXZ$ two-leg ladder model (\ref{ham1}) in the $\Delta$--$J$ plane
obtained from the groundstate fidelity and tensor network approach.
The various lines denote the phase boundaries between the nine different phases.
These are the ferromagnetic (FM),  stripe ferromagnetic (SF),  rung singlet (RS), rung triplet (RT),  N\'eel (N),
stripe N\'eel (SN), Haldane (H) and $XY$ ($XY1$ and $XY2$) phases.
Dashed lines denote a first-order phase transition and solid lines denote a continuous phase transition.} \label{phase}
\end{figure}

In this paper we use the TN algorithm developed in Ref.~\cite{lsh} to study a two-leg ladder model with
anisotropic spin-$\case12$ $XXZ$ interactions along the legs {and} rungs of the ladder.
A schematic phase diagram for this ladder model was mapped out previously using the sublattice entanglement of the groundstates~\cite{XXZ2}.
A variant of this ladder model, with isotropic rung interactions, was shown to exhibit a rich phase diagram~\cite{p1,XXZ1}.
We perform an extensive numerical analysis of the fully anisotropic model to map out the
groundstate phase diagram by evaluating the groundstate fidelity per
lattice site from the groundstate wave functions.
The phase transition points and thus the phase boundaries are detected through identifying pinch points on the groundstate fidelity surfaces,
which arise from the distinct changes of the groundstate wave functions in the vicinity of the critical
points as the model parameters are varied across a transition point.
This requires a scan of the entire parameter space.
The phase diagram unveiled in this way is shown in Figure~\ref{phase}.
Varying the $XXZ$ anisotropy parameter $\Delta$ and the relative rung coupling strength $J$ is seen to result in
a rich diversity of phases --
ferromagnetic (FM),  stripe ferromagnetic (SF),  rung singlet (RS), rung triplet (RT),  N\'eel (N),
stripe N\'eel (SN), Haldane (H) and two $XY$ phases ($XY1$ and $XY2$).
In particular, when $|\Delta| > 1$ the resulting Ising-type anisotropy breaks any continuous spin symmetry, in contrast to the $XY$-type regime for $|\Delta| < 1$.
The $XY$-type phases without any long-range order are characterized in terms of string order parameters.
In the symmetry-broken phases local order parameters are constructed from reduced density matrices for a representative groundstate wave function.
In the $XY$-type regime varying the rung coupling $J$ induces magnetic phases beyond standard ferromagnetism,
for example, the rung-singlet and Haldane phases.
For this model, we detect an additional rung-triplet phase as a result of the inherent competition between singlet formation and magnetic ordering.

The layout of the paper is as follows.
The fully anisotropic spin-$\case12$ $XXZ$ two-leg ladder model is described in Section~2, with the
results using the TN algorithm presented in detail in Section~3.
Concluding remarks, along with a discussion of previous results for this model, are given in Section~4.

\section{The model}

The fully anisotropic spin-$\case12$ $XXZ$ two-leg ladder system can be considered
as a pair of infinitely long spin chains coupled via rung interactions.
All spin interactions are of the spin-$\case12$ $XXZ$ Heisenberg type.
The Hamiltonian is defined by
\begin{equation}
H=H_{\mathrm{leg}}+H_{\mathrm{rung}},
\label{ham1}
\end{equation}
where
\begin{eqnarray}
  H_{\mathrm{leg}}&=&\sum_{i}\sum_{\alpha=1,2}
  \left(S_{\alpha, i}^{x}  \, S_{\alpha,i+1}^{x} + S_{\alpha, i}^{y} \, S_{\alpha, i+1}^{y}+\Delta \, S_{\alpha, i}^{z}  \, S_{\alpha,i+1}^{z}\right), \\
  H_{\mathrm{rung}}&=& J \sum_{i} \left( S_{1,i}^{x}  \, S_{2, i}^{x} + S_{1, i}^{y}  \, S_{2,i}^{y} + \Delta \, S_{1, i}^{z} \, S_{2,i}^{z} \right).
\end{eqnarray}
Here $S_{\alpha, i}^{\beta}$ ($\beta=x,y,z$) is the spin-$\case12$ operator acting at site $i$ on the $\alpha$-th leg,
$J$ is the rung coupling between the two spins on a rung, and $\Delta$ is the $XXZ$ anisotropy parameter.
Our aim is to map out the groundstate phase diagram of this model in the $\Delta$--$J$ plane.

In the FM phase, the ladder system orders ferromagnetically in both the leg and rung directions,
whereas in the SF phase it orders ferromagnetically in the leg direction and antiferromagnetically in the rung direction.
In contrast, in the N phase, the system orders antiferromagnetically in both the leg and rung directions,
whereas in the SN phase it orders antiferromagnetically in the leg direction and ferromagnetically in the rung direction.
Notice that in the FM, SF, N, and SN phases
$\langle S_{\alpha, i}^{x} \rangle= \langle S_{\alpha, i}^{y} \rangle =0$ and $\langle S_{\alpha,i}^{z} \rangle \neq0$.

The critical $XY$ phases belong to the universality class of the
Tomonaga-Luttinger liquid~\cite{TL}, which is known to exhibit a
power-law decay of the spin-spin correlations with gapless excitations.
The possibility for the existence of two different $XY$ phases in a ladder system
was pointed out in a slightly different model~\cite{p9}.
According to the Marshall-Lieb-Mattis theorem~\cite{MLM}, the $XY1$
phase is in the $J<0$ region and the $XY2$ phase is in the $J>0$ region of the parameter space.
Long-range order does not exist in the $XY$ phases in the thermodynamic limit.
Nevertheless, there is a practical way to characterize the $XY$ phases in the context of the
TN representation of the groundstate wave functions, which is achieved by defining a pseudo-order parameter
arising from the finiteness of the bond dimension $\chi$~\cite{whl}.
This offers a convenient means to numerically determine the phase boundaries
between the $XY$ phase and other phases.
For the $XY1$  and $XY2$ phases, $\langle S_{\alpha,i}^{z} \rangle =0$
and $\langle S_{\alpha, i}^{x} \rangle^2 + \langle S_{\alpha, i}^{y} \rangle^2 \neq0$.

The two-spin states on a rung are written as
\begin{eqnarray*}
|0,0\rangle&=&{1\over\sqrt2}(| \! \uparrow_1 \downarrow_2\rangle-| \! \downarrow_1 \uparrow_2\rangle),\\
|1,x\rangle&=&{1\over\sqrt2}(| \! \downarrow_1 \downarrow_2\rangle-| \! \uparrow_1\uparrow_2\rangle),\\
|1,y\rangle&=&{{\mathrm{i}}\over\sqrt2} (| \! \downarrow_1\downarrow_2 \rangle +| \! \uparrow_1 \uparrow_2\rangle,\\
|1,z\rangle&=&{1\over\sqrt2} (| \! \uparrow_1 \downarrow_2\rangle+| \! \downarrow_1 \uparrow_2\rangle).
\end{eqnarray*}
Here the subscripts $1$ and $2$ label the different spins on the same rung (see, e.g., Ref.~\cite{XY12}).
The first state is the singlet and the remaining three states constitute the triplet.
As can readily be seen from the Hamiltonian (\ref{ham1}), the groundstate of the system in the strong coupling limit $J\to\pm\infty$
corresponds to the RS ($|0,0\rangle$) and RT ($|1,z\rangle$) phases, respectively.
It should be noted that the RS phase, the RT phase and the H phase lack long-range order in the conventional
Landau-Ginzburg-Wilson sense.
However, each are characterized by a suitably modified string order parameter~\cite{string}.
For the RS, RT  and H phases $\langle S_{\alpha, i}^{\beta} \rangle =0$.

\section{Results}

To map out the groundstate phase diagram, we first consider the regions $J>0$ and $J<0$, with
the anisotropy $\Delta$ as a control parameter.
We then fix $\Delta$ and vary the rung coupling $J$.
Illustrative examples are given in detail below for the values $J=\pm1$.
In our approach, apart from order parameters, which we also make use of,
the groundstate fidelity per lattice site is used as an important tool to detect quantum phase transitions.
For two different groundstates $|\psi(x)\rangle$ and $|\psi(x')\rangle$ in a quantum system corresponding to different values $x$ and $x'$
of a control parameter,
the fidelity $F(x,x')=|\langle\psi(x)|\psi(x')\rangle|$ is defined as a measure of the overlap between the two states.
For large but finite size $N$, the fidelity $F$ scales as $d^N$,
with $d$ interpreted as the ground state fidelity per site  $d=\lim_{N\rightarrow\infty}{F_N^{1/N}}$.
As a well-defined property even in the thermodynamic limit, the groundstate fidelity per lattice site $d$
enjoys some properties inherited from the fidelity $F$, namely:
(i) symmetry under interchange, $d(x,x')=d(x',x)$;
(ii) normalization, $d(x,x)=1$;
(iii) range $0\leq d(x,x')\leq 1$.

In previous studies the fidelity per lattice site $d$ was demonstrated to be a useful detector for
different types of quantum phase transition via its singularity structure~\cite{zhou,zhou1,b6,b7,b3,whl}.
In this approach, a pinch point in the $d(x,x')$ surface occurs at a quantum phase transition,
i.e., at each phase transition point $x_c$, a pinch point $(x_c,x_c)$ occurs on the surface of fidelity per lattice site $d(x,x')$
at the crossing of two singular lines $x=x_c$ and $x'=x_c$.
In these studies the groundstate wavefunctions are determined using TN algorithms~\cite{peps,TNreview}.
The combined groundstate fidelity and TN approach has been tailored to include ladder systems~\cite{lsh}.

A key point in the investigation of the two-leg quantum spin ladders via the TN approach
is that the computational cost scales as $\chi^6$, where $\chi$ is the bond dimension
for each of the four necessary four-index tensors.
These tensors need to be updated simultaneously, with the
updating procedure closely related to the infinite PEPS~\cite{PEPS} and
translationally invariant MPS~\cite{MPS} algorithms.
A summary of the TN approach for quantum spin ladders is given in Appendix A.
For the spin-$\case12$ isotropic $XXX$ two-leg ladder it was demonstrated earlier that the TN approach
could accurately determine the groundstate wave function, with bond dimension $\chi=6$
outperforming the DMRG results~\cite{lsh}.
Of necessity for the calculation of the order parameters, the reduced density matrix $\rho$ can be computed directly
from the TN representation of the groundstate wave functions~\cite{lsh}.
The reduced density matrix displays different nonzero-entry structures in different phases.

\begin{figure}
\begin{center}
\includegraphics[width=0.6\textwidth]{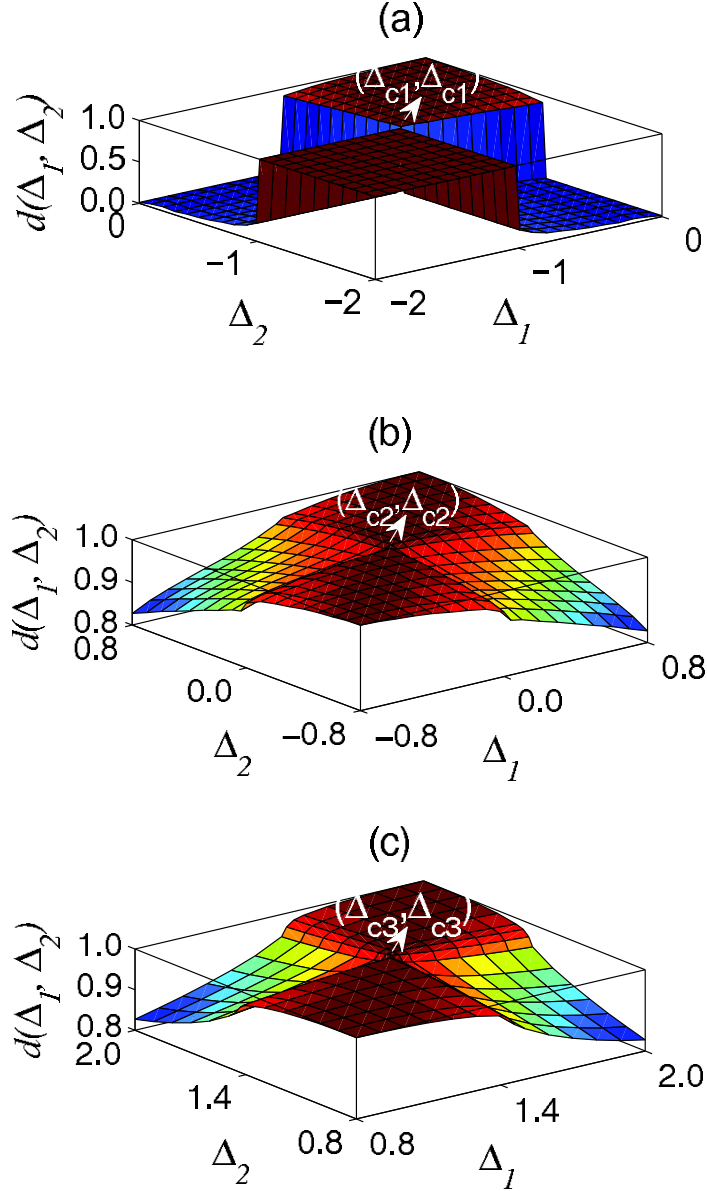}
\end{center}
\caption{The groundstate fidelity surface $d(\Delta_1,\Delta_2)$ for
the spin-$\case12$ $XXZ$ two-leg ladder model (\ref{ham1}) for rung coupling $J=1$
and varying anisotropy $\Delta$ calculated with bond dimension $\chi=6$.
Three pinch points are identified at $(\Delta_{c1},\Delta_{c1})$,
$(\Delta_{c2},\Delta_{c2})$ and $(\Delta_{c3},\Delta_{c3})$ on the global
fidelity surface, indicating three phase transition
points, located at (a) $\Delta_{c1}= -1.00$, (b) $\Delta_{c2}= 0.00$
and (c) $\Delta_{c3}= 1.43$.}
\label{DDD1}
\end{figure}

\subsection{$J=1, -2 \le \Delta \le 2$}

\begin{figure}
\begin{center}
\includegraphics[width=0.6\textwidth]{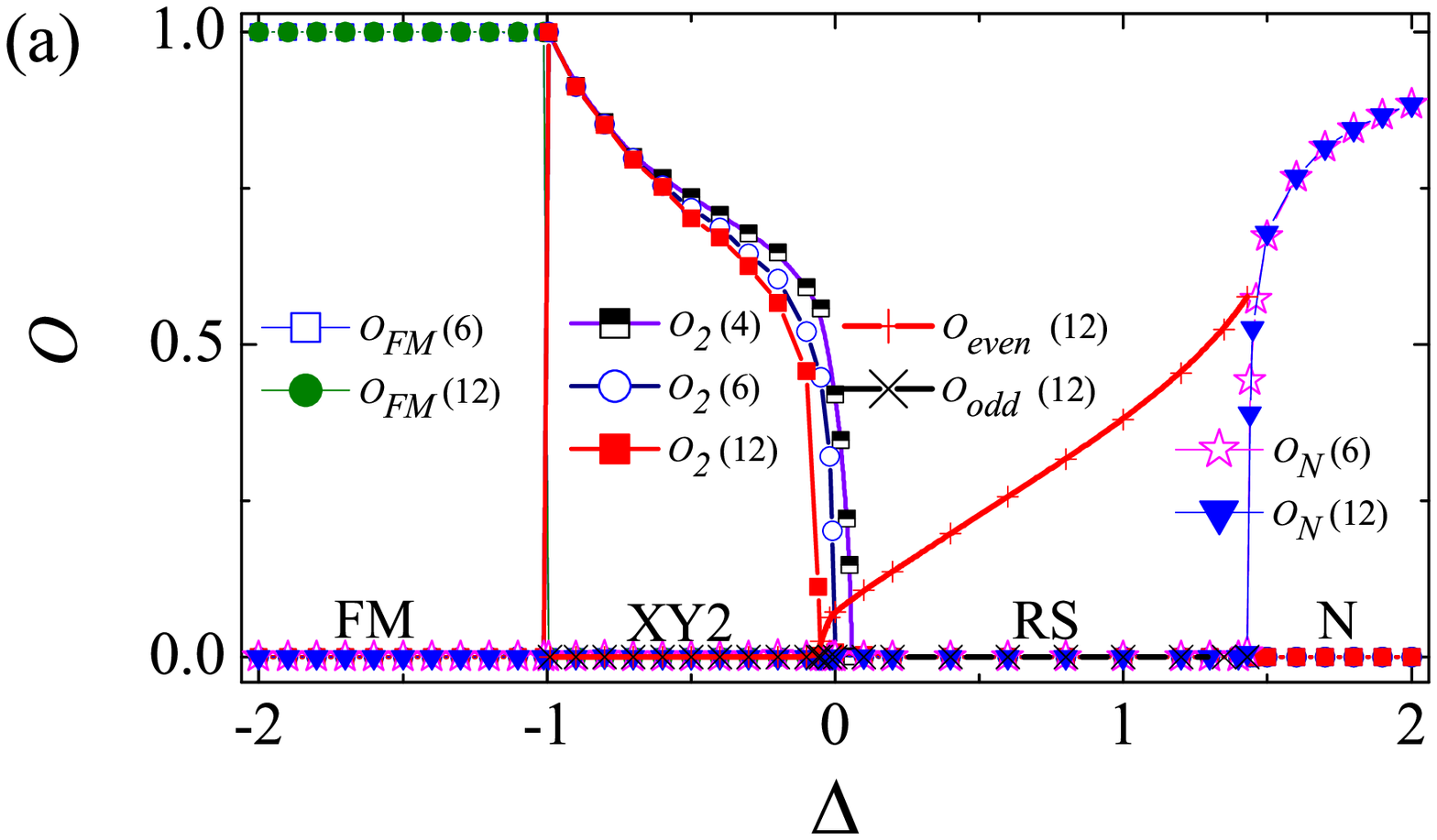}
\includegraphics[width=0.6\textwidth]{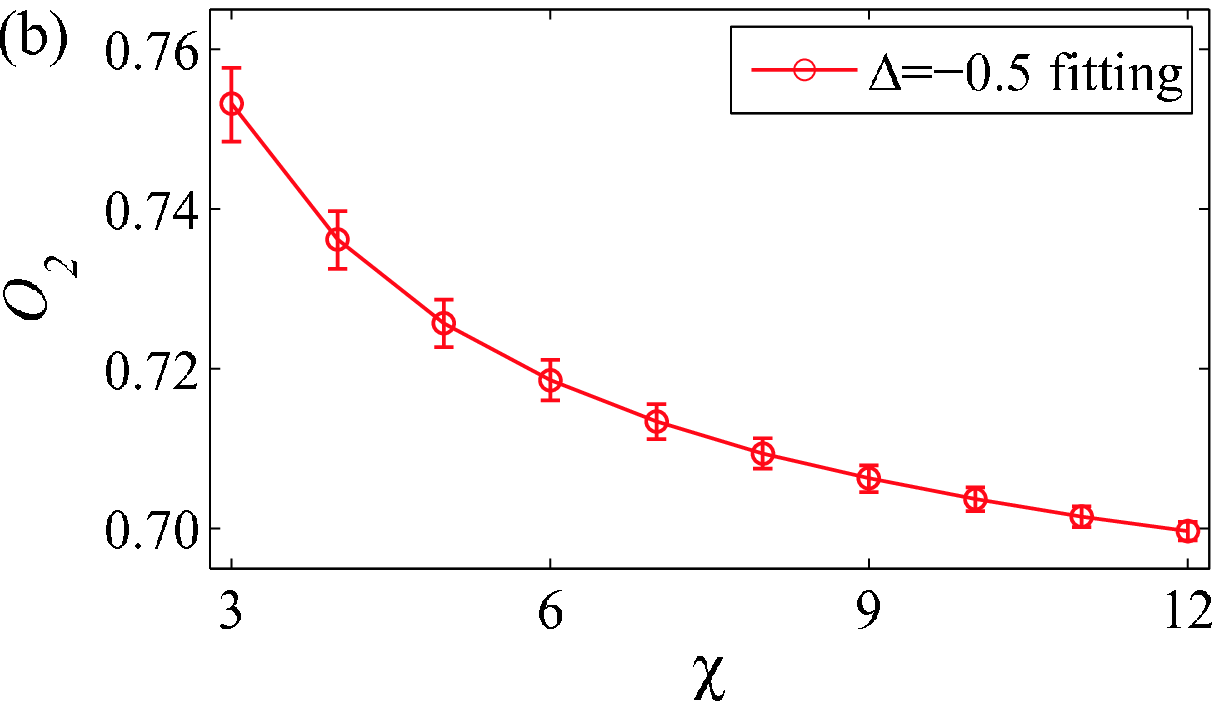}
\includegraphics[width=0.6\textwidth]{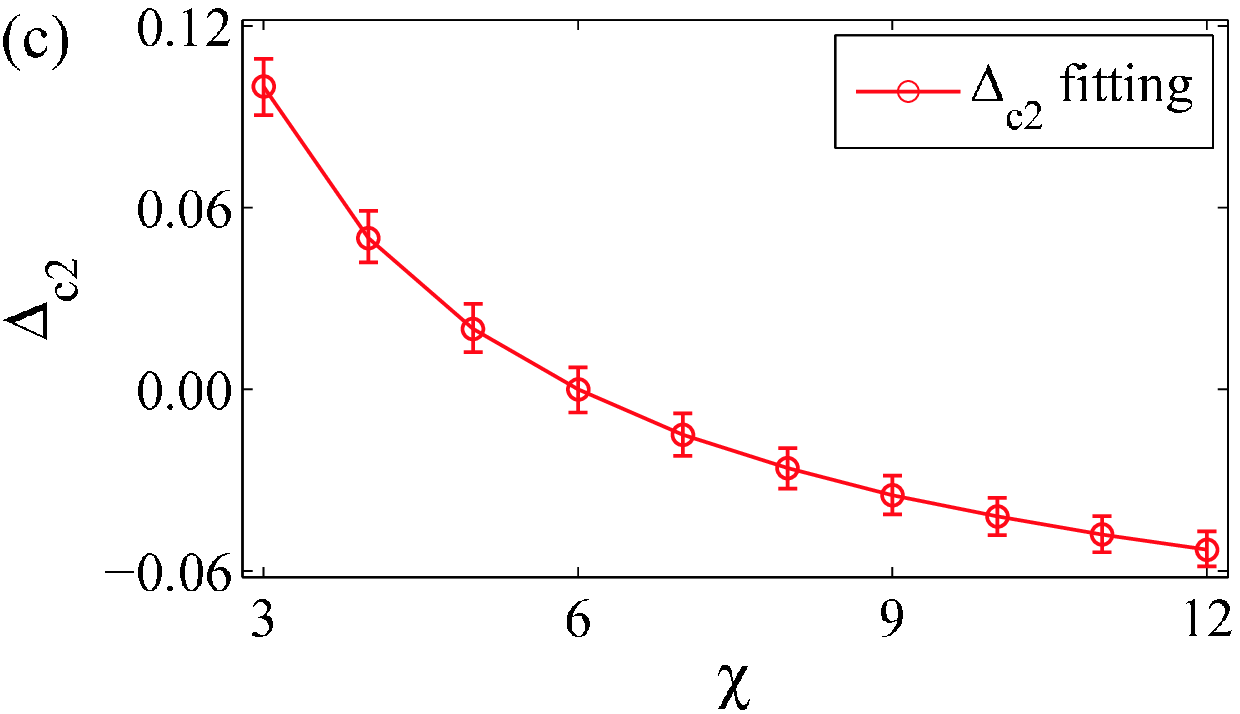}
\end{center}
\caption{(a) The local order parameter
$O_{\mathrm{FM}}(\chi)$ in the FM phase, the pseudo-order parameter $O_{2}(\chi)$ in the
$XY2$ phase, the local order parameter $O_{\mathrm{N}}(\chi)$ in the N\'eel
phase and the string order parameters $O_{\mathrm{odd}}(\chi)$ and $O_{\mathrm{even}}(\chi)$ in the RS
phase as a function of the anisotropy $\Delta$, with $J=1$.
Three phase transition points, $\Delta_{c1}$, $\Delta_{c2}$ and $\Delta_{c3}$, are distinguished by the
behavior of the order parameters.
The phase transition points $\Delta_{c1}=-1.00$ and $\Delta_{c3}=1.43$ are saturated for bond dimension
$\chi=6$, whereas $\Delta_{c2}$ shifts with increasing $\chi$.
(b) The scaling of the pseudo-order parameter $O_{2}(\chi)$ for $\Delta=-0.5$ (see text).
(c) The ``pseudo-critical" point estimates $\Delta_{c2}(\chi)$ obtained
from the pseudo-order parameter $O_{2}(\chi)$ and their fitting function (see text).
 }\label{OR1}
\end{figure}

Fixing the rung coupling to the value $J=1$ and varying the anisotropy parameter $\Delta$ from $-2$ to $2$,
we are able to determine the phase boundaries between the FM phase, the $XY2$ phase, the RS phase and the N\'eel phase.
The groundstate fidelity per site, $d(\Delta_1,\Delta_2)$, is shown in Figure~\ref{DDD1} as a function of $\Delta_1$ and $\Delta_2$,
calculated with bond dimension $\chi=6$.
Here the range of the plots is restricted to highlight the fidelity surface and, in particular, the pinch points.
The results in Figure~\ref{DDD1}(a) and Figure~\ref{DDD1}(c) differ very little for a larger bond dimension $\chi$,
indicating that the computational results are almost saturated for bond dimension $\chi=6$.
However, the pinch point in Figure~\ref{DDD1}(b) shifts noticeably when $\chi$ increases
from $\chi=6$ to $\chi=12$, as Figure~\ref{OR1} shows.
Three pinch points are identified at $(\Delta_{c1}, \Delta_{c1})$,
$(\Delta_{c2}, \Delta_{c2})$ and $(\Delta_{c3}, \Delta_{c3})$ on the
fidelity surfaces, indicating that there are three phase transition points.
These are located at (i) $\Delta_{c1}= -1.00$, (ii) $\Delta_{c2}= 0.00$ and (iii) $\Delta_{c3}= 1.43$.
In this way we are able to identify four different phases, separated by three transition points.
It remains to characterize these phases in terms of local or nonlocal order parameters.

In the FM phase, the one-site reduced density matrix yields the local order parameter,
\begin{equation}\label {fitlader1}
O_{\mathrm{FM}}=\frac{1}{2} |\langle (S^z_{1, i}+ S^z_{2, i})+ (S^z_{1,i+1}+S^z_{2, i+1}) \rangle|.
\end{equation}
Our numerical results indicate that the ladder system is in the FM
phase for $\Delta < -1.00$ (see Figure~\ref{OR1}(a)).

In the N\'eel phase, we similarly consider the local order parameter
\begin{equation}\label {fitlader2}
O_{\mathrm{N}}=\frac{1}{2} |\langle (S^z_{1, i}- S^z_{2, i})- (S^z_{1,i+1}-S^z_{2, i+1}) \rangle |.
\end{equation}
In the N\'eel phase, the N\'eel order $O_{\mathrm{N}}$ exhibits a long-range order.
The evaluation of the order parameter indicates that the ladder system is in the N\'eel
phase for $\Delta > 1.43$ (see Figure~\ref{OR1}(a)).

\begin{figure}
\begin{center}
\includegraphics[width=0.48\textwidth]{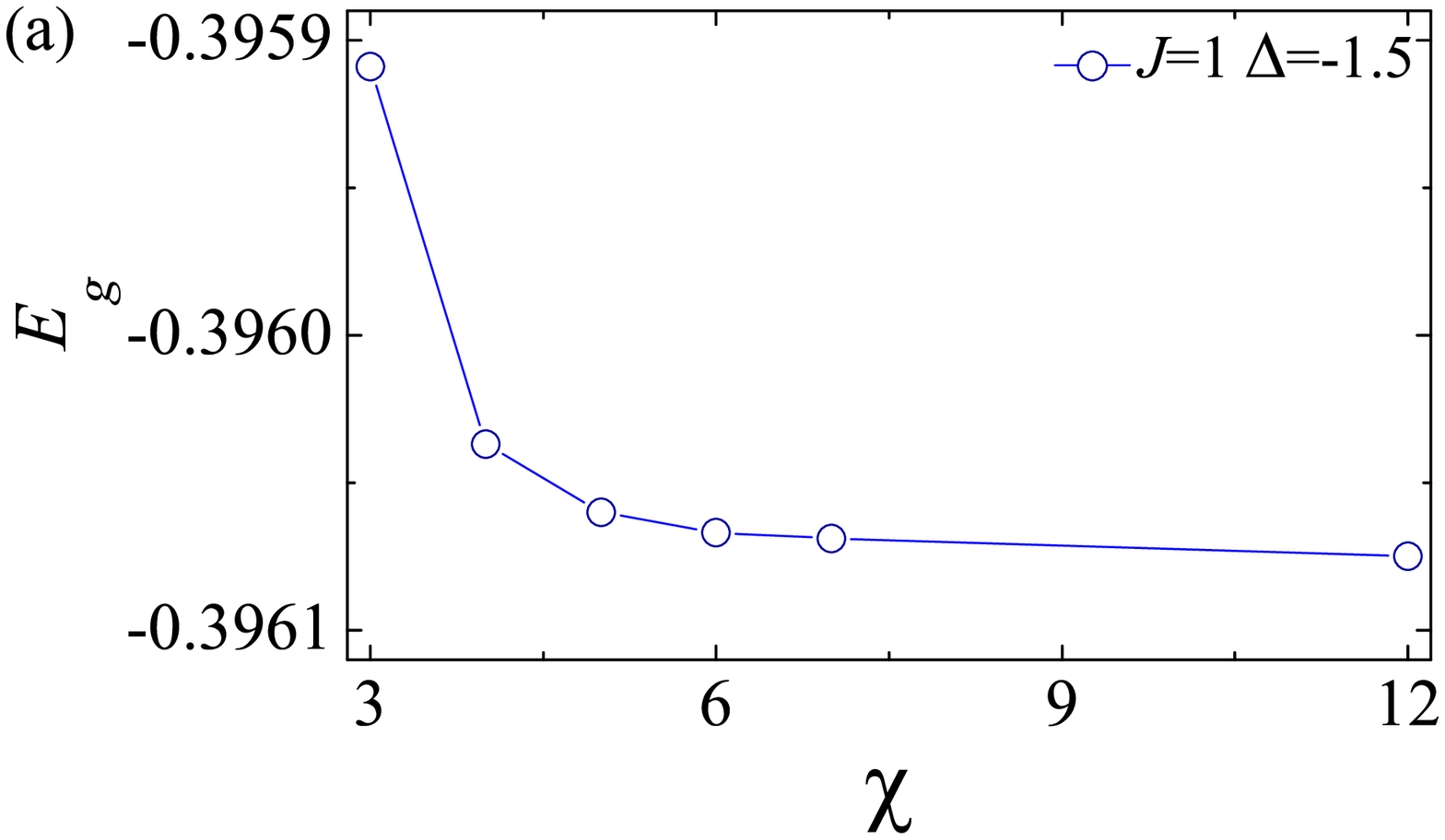}
\hspace{2mm}
\includegraphics[width=0.48\textwidth]{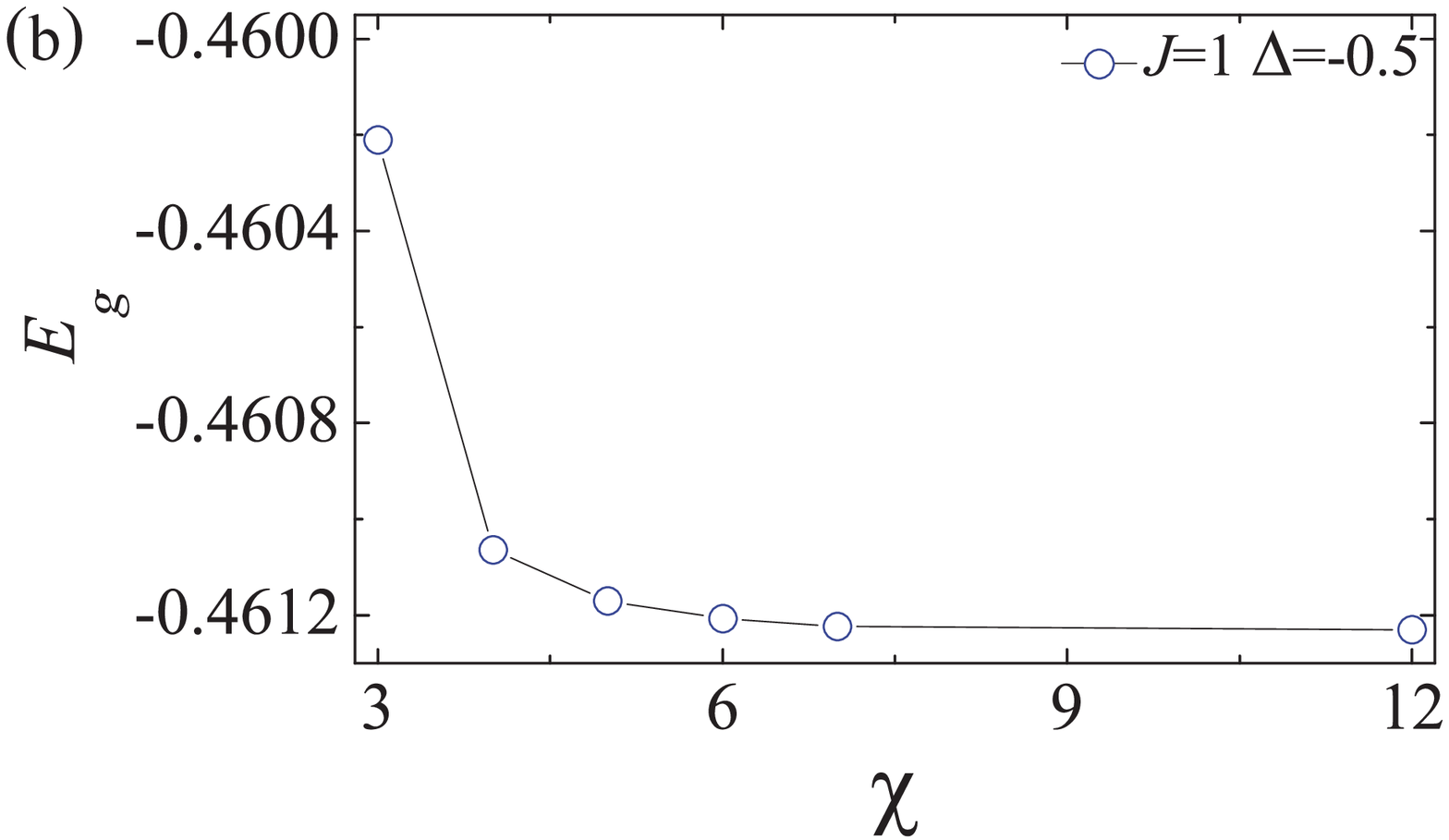}\\
\vspace{4mm}
\includegraphics[width=0.48\textwidth]{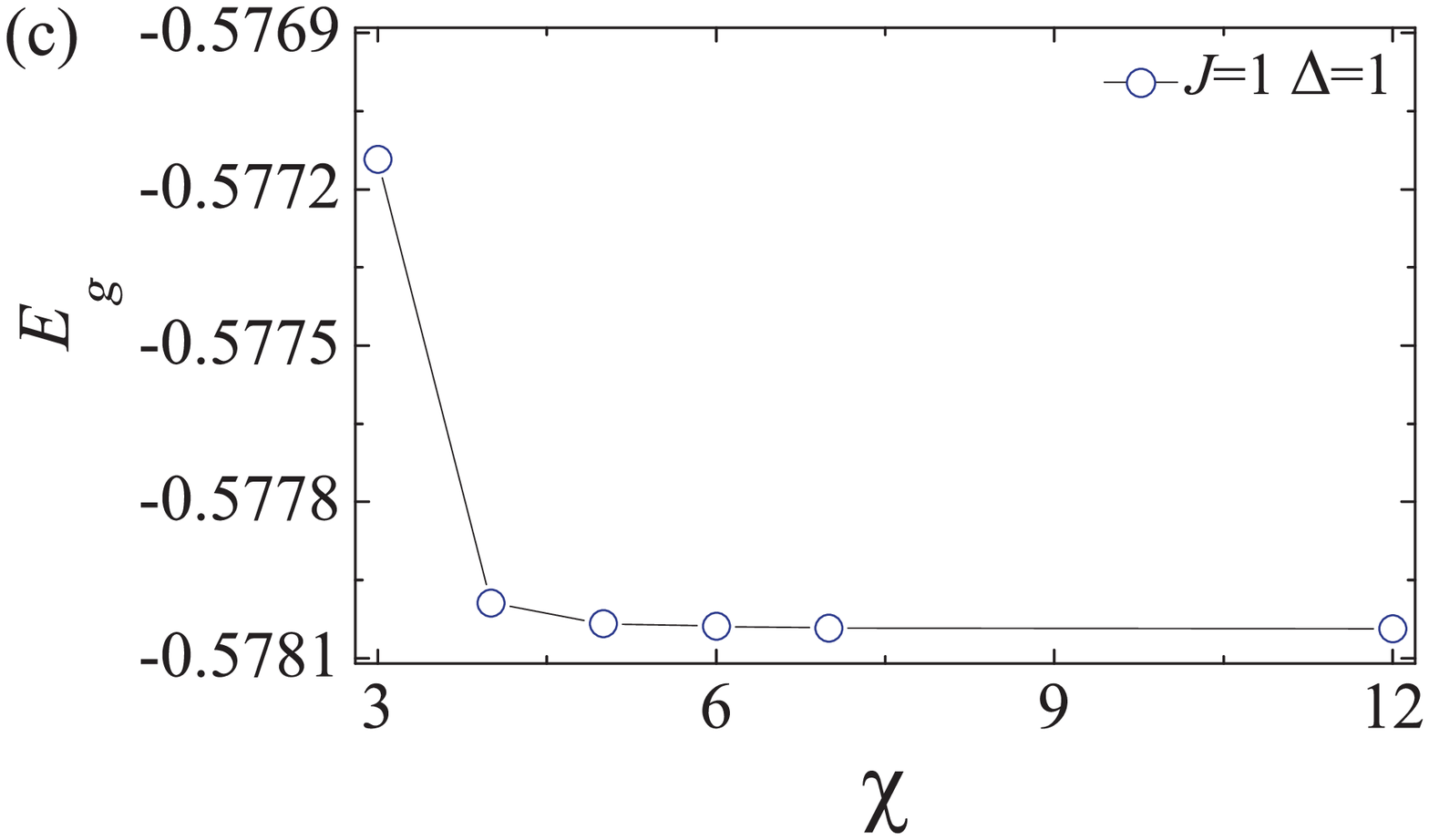}
\hspace{2mm}
\includegraphics[width=0.48\textwidth]{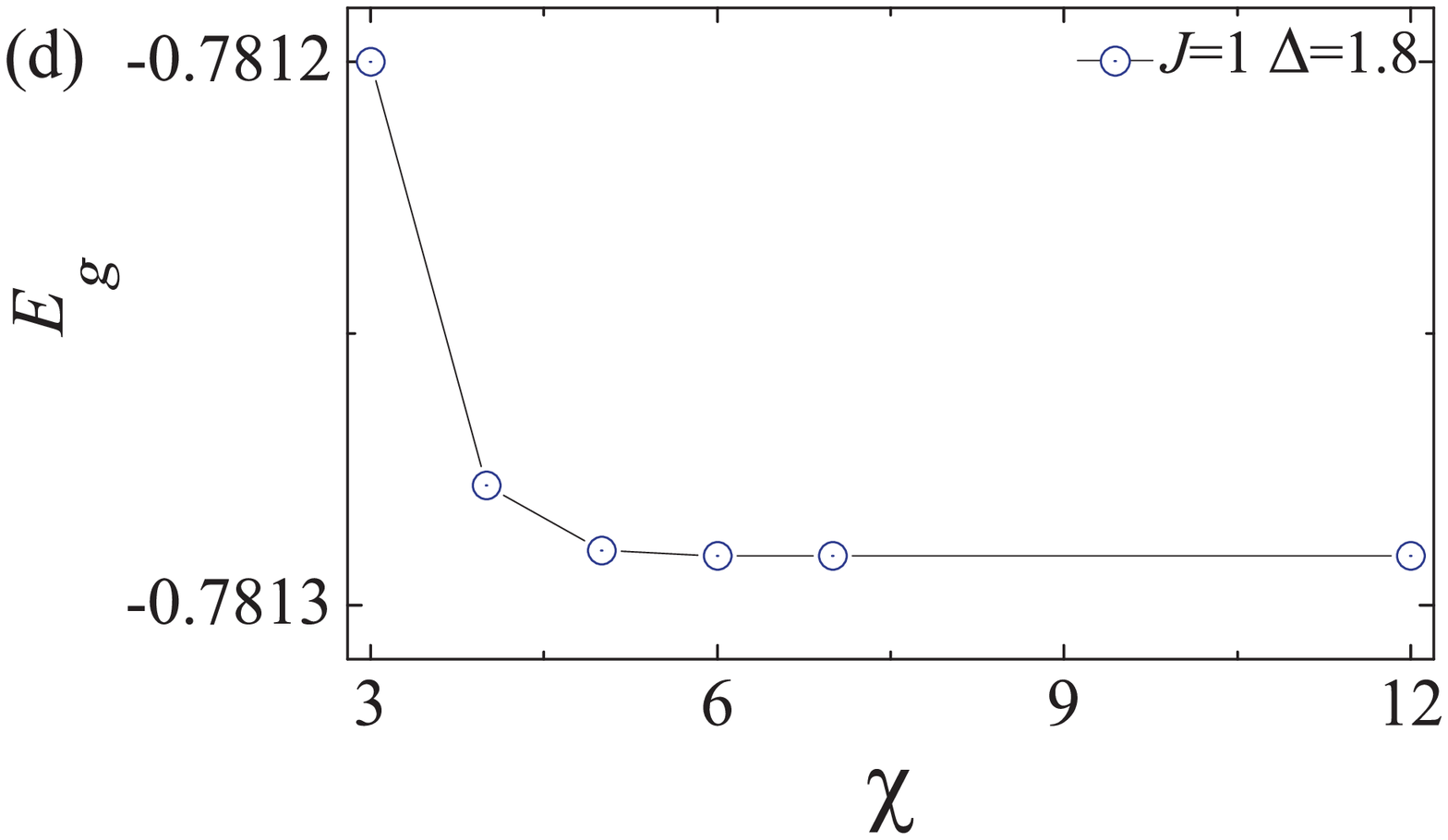}
\end{center}
\caption{Convergence of the groundstate energy $E_g$ with increasing bond dimension $\chi=3,4,5,6,7,12$ in different phases
with rung coupling $J=1$ and anisotropy parameter values
(a) $\Delta=-1.5$ (FM phase)
(b) $\Delta=-0.5$ ($XY2$ phase)
(c) $\Delta=1.0$ (RS phase)
(d) $\Delta=1.8$ (N phase).
}\label{Eg1}
\end{figure}

\begin{figure}
\begin{center}
\includegraphics[width=0.6\textwidth]{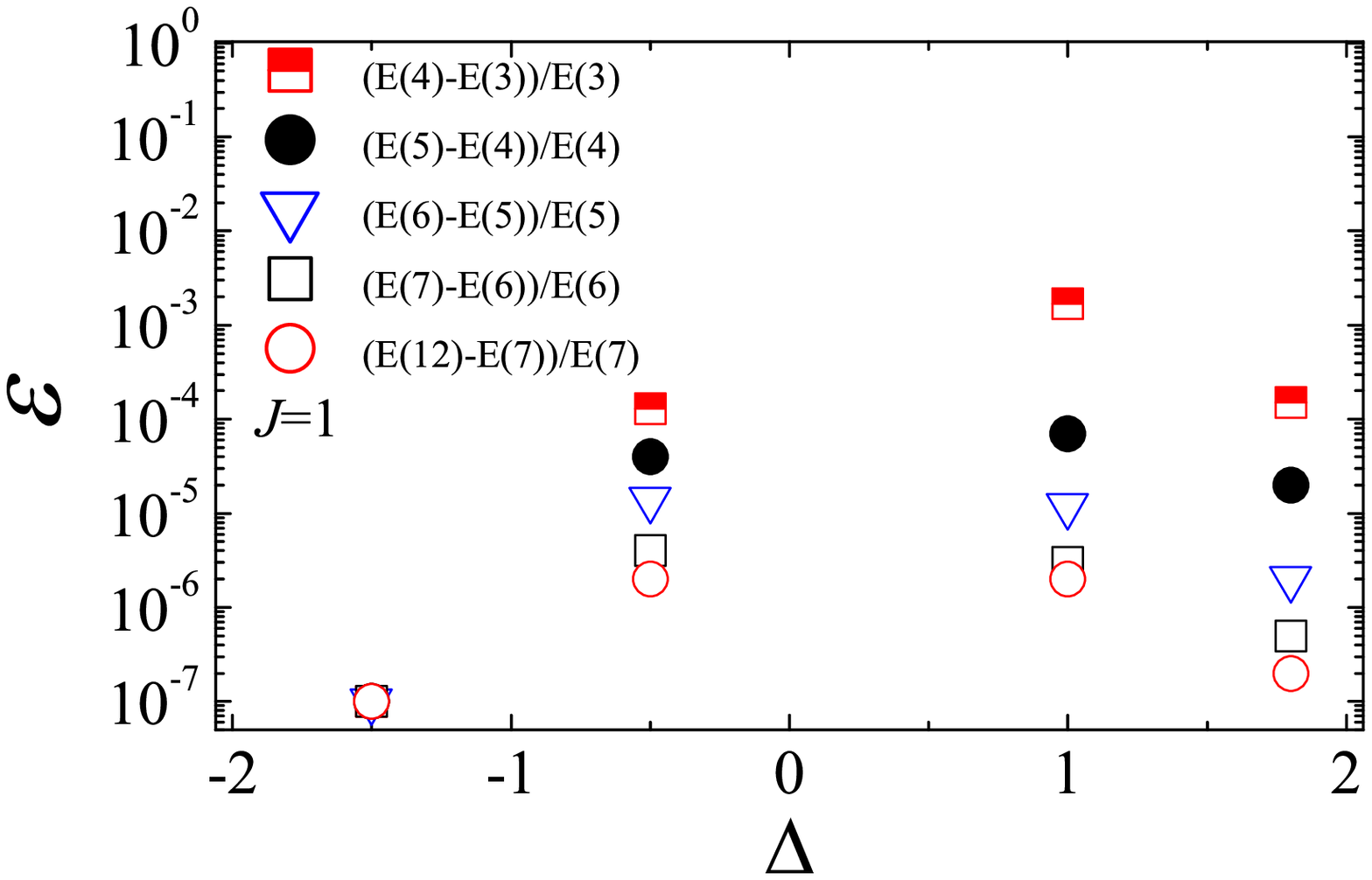}
\end{center}
\caption{Relative error $\varepsilon$ in the estimates of the groundstate energy as a function of $\Delta$ at $J=1$ for increasing bond dimension $\chi$.
The relative errors are defined by $\varepsilon = (E(\chi_{n+1})-E(\chi_n))/E(\chi_n)$ where $E(\chi_n)$
is the groundstate energy estimate for given bond dimension $\chi_n$.
 }\label{EEE1}
\end{figure}

\begin{figure}
\begin{center}
\includegraphics[width=0.48\textwidth]{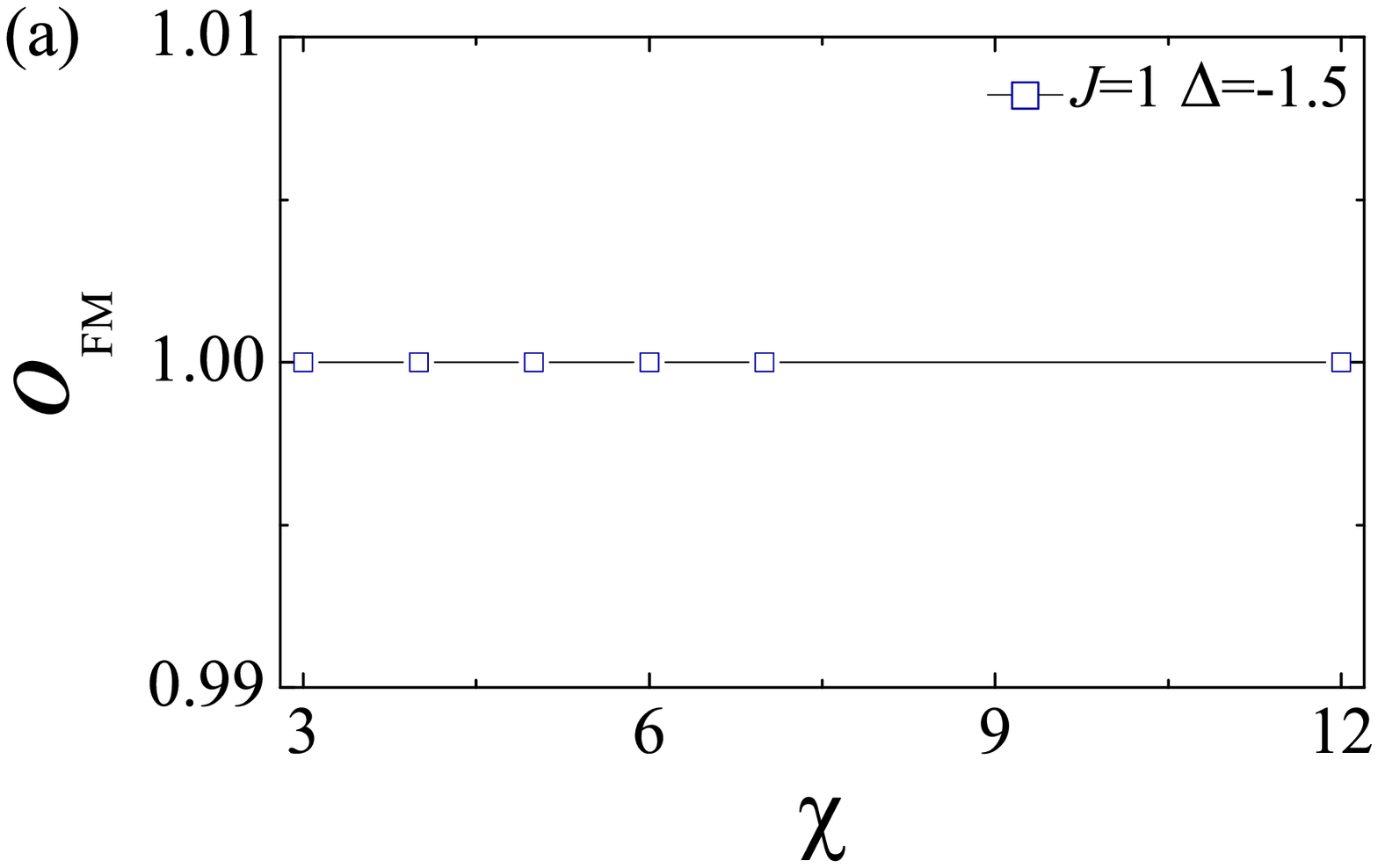}
\hspace{2mm}
\includegraphics[width=0.48\textwidth]{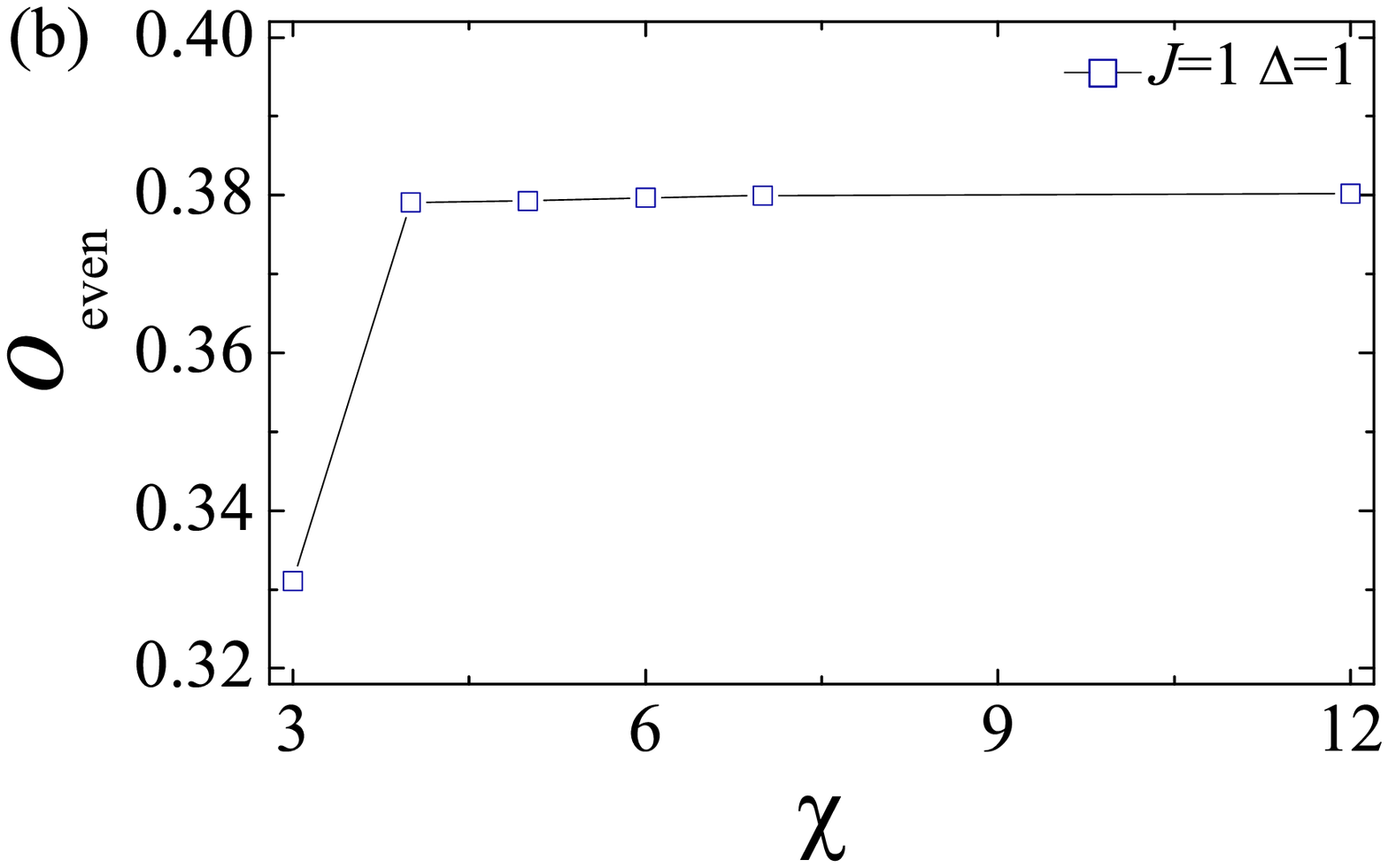}\\
\vspace{4mm}
\includegraphics[width=0.48\textwidth]{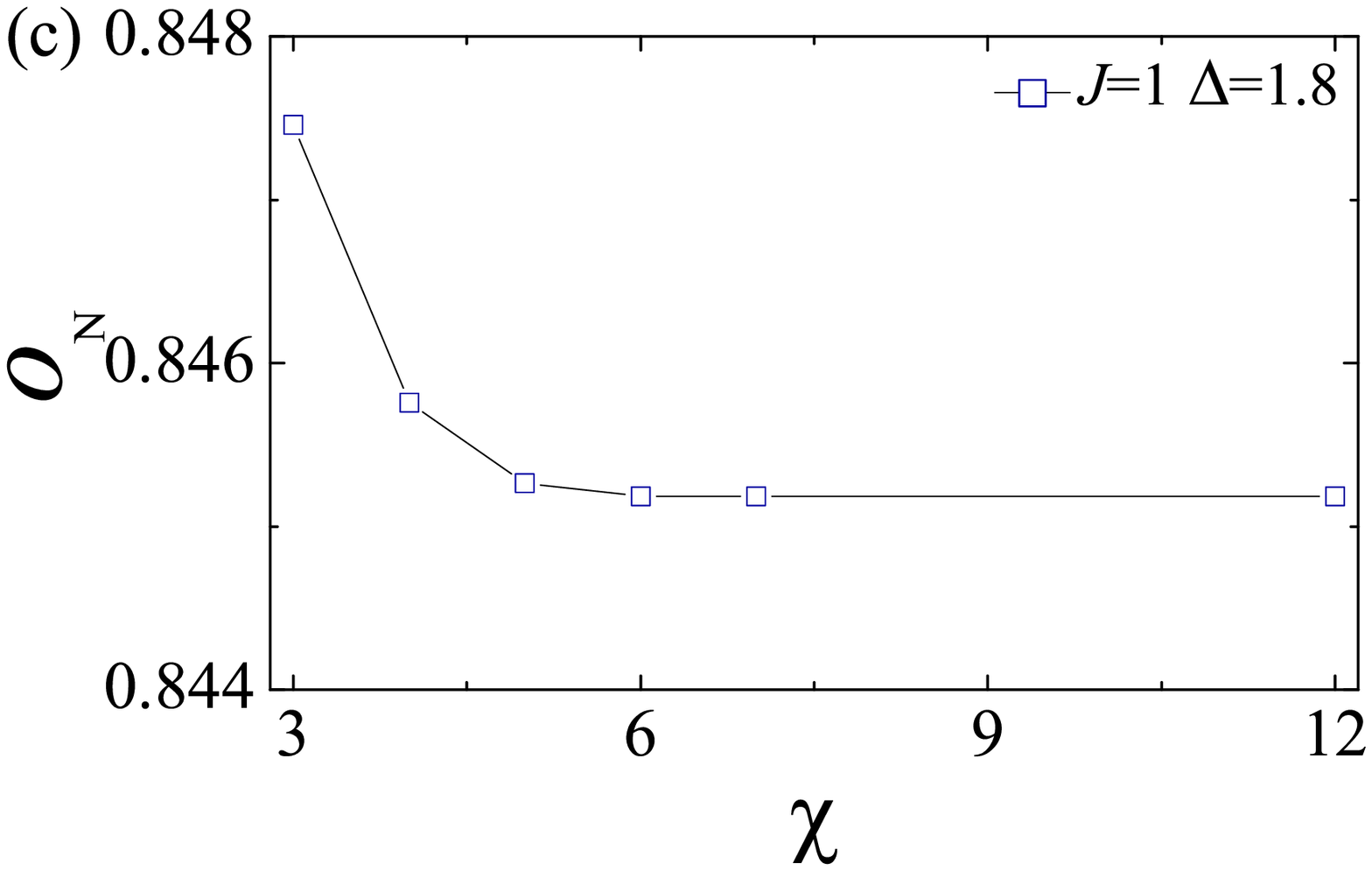}
\end{center}
\caption{Convergence of order parameters with increasing bond dimension $\chi=3,4,5,6,7,12$ in different phases for $J=1$.
(a) local order parameter $O_{\mathrm{FM}}$ for $\Delta=-1.5$ (FM phase).
(b) string order parameter $O_{\mathrm{even}}$ for $\Delta=1.0$ (RS phase).
(c) local order parameter $O_{\mathrm{N}}$ for $\Delta=1.8$ (N phase).
 }\label{O1}
\end{figure}

In the RS phase, the groundstate wave function may be approximated by the product of local rung singlets.
As usual, two distinct string order parameters, $O_{\mathrm{odd}}$ and $O_{\mathrm{even}}$,
may be used to characterize the RS phase.
For two-leg ladder models these are defined by
\begin{equation}
O_{{\mathrm{odd}}/{\mathrm{even}}} = - \lim_{ \left|i-j\right| \to \infty}
\Big\langle S_{{\mathrm{o}}/{\mathrm{e}},i}^z
\exp \Big[ \mathrm{i} \pi \sum_{l=i+1}^{j-1} S_{{\mathrm{o}}/{\mathrm{e}},l}^z \Big] S_{{\mathrm{o}}/{\mathrm{e}},j}^z
 \Big\rangle,
\label{stringorder}
\end{equation}
where $S^z_{{\mathrm{o}},i} \equiv S^z_{1,i}+S^z_{2,i}$ and $S^z_{{\mathrm{e}},i} \equiv
S^z_{1,i}+S^z_{2,i+1}$.
The odd and even string order parameters are actually mutually exclusive in the RS, RT and H phases.
Our results for $O_\mathrm{odd}$ and $O_\mathrm{even}$ indicate that
the RS phase persists in the region $\Delta_{c2} < \Delta < \Delta_{c3}$ with $J=1$ (see Figure~\ref{OR1}(a)).

In the $XY$ phases,  the TN algorithm automatically leads to infinite degenerate
groundstates, arising from pseudo spontaneous symmetry  breaking of the
continuous U(1) symmetry, due to the finiteness of the bond dimension $\chi$.
This allows the introduction of a pseudo-order parameter that must scale to zero,
in order to keep consistency with the Mermin-Wagner theorem.
As suggested in Refs.~\cite{whl} and~\cite{XY12},
the pseudo-order parameters in the $XY1$ and $XY2$ phases may be defined as
\begin{equation}\label
{fitlader32}
O_{1}=\sqrt{\langle  S^{x}_{1, i}+ S^{x}_{2, i}  \rangle ^2 +
\langle S^{y}_{1, i}+ S^{y}_{2, i}  \rangle ^2},
\end{equation}
and
\begin{equation}\label
{fitlader31}
O_{2}=\sqrt{\langle  S^{x}_{1, i}- S^{x}_{2, i}  \rangle ^2 +
\langle S^{y}_{1, i}- S^{y}_{2, i}  \rangle ^2}.
\end{equation}

The pseudo-order parameter $O_{2}$ in the $XY2$ phase is plotted as a function of
the anisotropy $\Delta$ for different values of the bond dimension $\chi$ in Figure~\ref{OR1}(a).
This order parameter $O_{2}$ is also plotted as a
function of $\chi$ for $\Delta=-0.5$ in Figure~\ref{OR1}(b).
It is clearly seen that $O_{2}$ decreases as $\chi$ increases.
Here the pseudo-order parameter scales to zero according to
$O_2(\chi)=a_1\, \chi^{-b_1}(1+c_1\, \chi^{-1})$, with $a_1=0.697(3)$, $b_1=0.007(1)$ and $c_1=0.26(1)$.
Such scaling was adopted in previous studies of pseudo-order parameters~\cite{whl,XY12} and ensures
that they vanish for infinite bond dimension as expected.

The estimates of the ``pseudo-critical'' values $\Delta_{c2}(\chi)$ at which the pseudo-order parameter $O_{2}(\chi)$ becomes
zero are shown in Figure~\ref{OR1}(c).
These points are extrapolated with respect to the bond dimension $\chi$, with
an extrapolation function $\Delta(\chi)=a_2+b_2 \, \chi^{-c_2}$.
The results imply the fitting constants $a_2=-0.110(2)$, $b_2=0.585(7)$ and $c_2=0.93(2)$.
In the limit $\chi\rightarrow\infty$ we have $\Delta(\infty)=a_2$, which
is the estimate for the critical point $\Delta_{c2}$  between the $XY2$ and RS phases.
The term ``pseudo-critical" point is used because such estimates are obtained using the pseudo-order parameters.
Nevertheless they yield estimates for real critical points.

The existence of the FM phase, the $XY2$ phase, the N\'eel phase
and the RS phase is numerically confirmed in this way, as seen from  Figure~\ref{OR1}.
Specifically, the three phase transition points occur at the values $\Delta_{c1}=-1.00$, $\Delta_{c2}\simeq-0.11$
and $\Delta_{c3}=1.43$.
If the anisotropy coupling $\Delta$ is tuned as a control parameter, the ladder system undergoes a first-order phase
transition at $\Delta_{c1}$, with continuous phase transitions at $\Delta_{c2}$ and $\Delta_{c3}$.
The different nature of the first-order transition at $\Delta_{c1}$
can be seen clearly in the fidelity surface in Figure~\ref{DDD1}(a), compared to the
continuous transitions in Figure~\ref{DDD1}(b) and Figure~\ref{DDD1}(c).
In general, for continuous phase transitions, the fidelity surface shows continuous behaviour near critical points,
while discontinuous phase transitions show discontinuous behaviour.

In the limit $J\rightarrow\infty$ the $XY2$ phase vanishes.
In addition for general $J>0$ the line $\Delta =-1$ is the phase boundary between the FM phase
$\Delta <-1$ and the $XY2$ phase $\Delta >-1$.
We have also confirmed that $\Delta \approx J$ is the phase boundary between
the RS phase and the N\'eel phase, when $\Delta\rightarrow \infty$ and $J\rightarrow\infty$.
There are thus three phases when $J\rightarrow\infty$:  the FM phase, the RS
phase and the N\'eel phase.

Concerning the accuracy of our results, estimates of the groundstate energy per site with increasing bond dimension
are shown in Figure~\ref{Eg1}.
The estimates are seen to converge rapidly, as quantified in Figure~\ref{EEE1}.
We find that almost all of the relative errors reach to order $10^{-5}$.
This indicates the reliability of the algorithm for generating accurate groundstate wave functions for spin ladders,
given a preset error tolerance.
The corresponding convergence of the order parameter estimates is shown in Figure~\ref{O1}.
As is well understood, states in gapless phases, such as the $XY1$ and $XY2$ phases, are more difficult to compute than states in
phases with a large gap, such as the FM, N and SF phases, which converge much faster, yielding the same accuracy with a smaller bond dimension.
Phases with a small gap, such as the H phase, are also more difficult to compute.
In general this slower convergence in gapless (critical) versus gapped phases is a feature of numerical entanglement
based approaches, such as TN methods~\cite{TNreview}.
This is because the overall strategy of the TN method is more efficient for gapped systems.
By design, the finite bond dimension already induces a gap in the system.
From the perspective of entanglement in quantum critical phenomena~\cite{entangle},
gapped systems obey an area law, which can be well described using relatively small bond dimensions.
On the other hand, for gapless systems, the entanglement diverges as the gap vanishes, requiring larger, in principle infinite, bond dimensions.
Alternatively, for the same size bond dimension, more iteration steps are necessary for convergence in gapless compared to gapped systems.
Our algorithm is seen to work sufficiently well in the phases under consideration.
However, the critical $XY1$ and $XY2$ phases require additional extrapolation to infinite-size bond dimension.

\begin{figure}
\begin{center}
\includegraphics[width=0.98\textwidth]{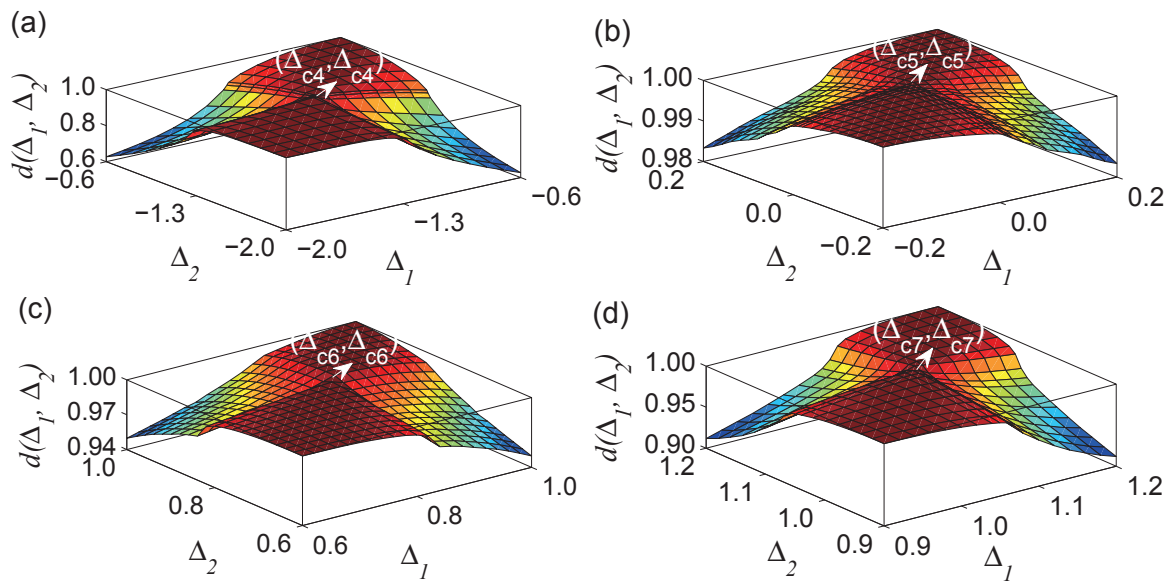}
\end{center}
\caption{The groundstate fidelity surface $d(\Delta_1,\Delta_2)$ for
the spin-$\case12$ $XXZ$ two-leg ladder model (\ref{ham1}) for rung coupling $J=-1$
and varying anisotropy $\Delta$ calculated with bond dimension $\chi=6$.
Four pinch points are identified at $(\Delta_{c4},\Delta_{c4})$, $(\Delta_{c5},\Delta_{c5})$,
$(\Delta_{c6},\Delta_{c6})$ and $(\Delta_{c7},\Delta_{c7})$ on the global
fidelity surface, indicating four phase transition
points, located at (a) $\Delta_{c4}= -1.26$, (b) $\Delta_{c5}= 0.00$, (c) $\Delta_{c6}=
0.84$, and (d) $\Delta_{c7}= 1.06$.
} \label{DDD2}
\end{figure}

\subsection{$J=-1, -2 \le \Delta \le 2$}

\begin{figure}
\begin{center}
\includegraphics[width=0.49\textwidth]{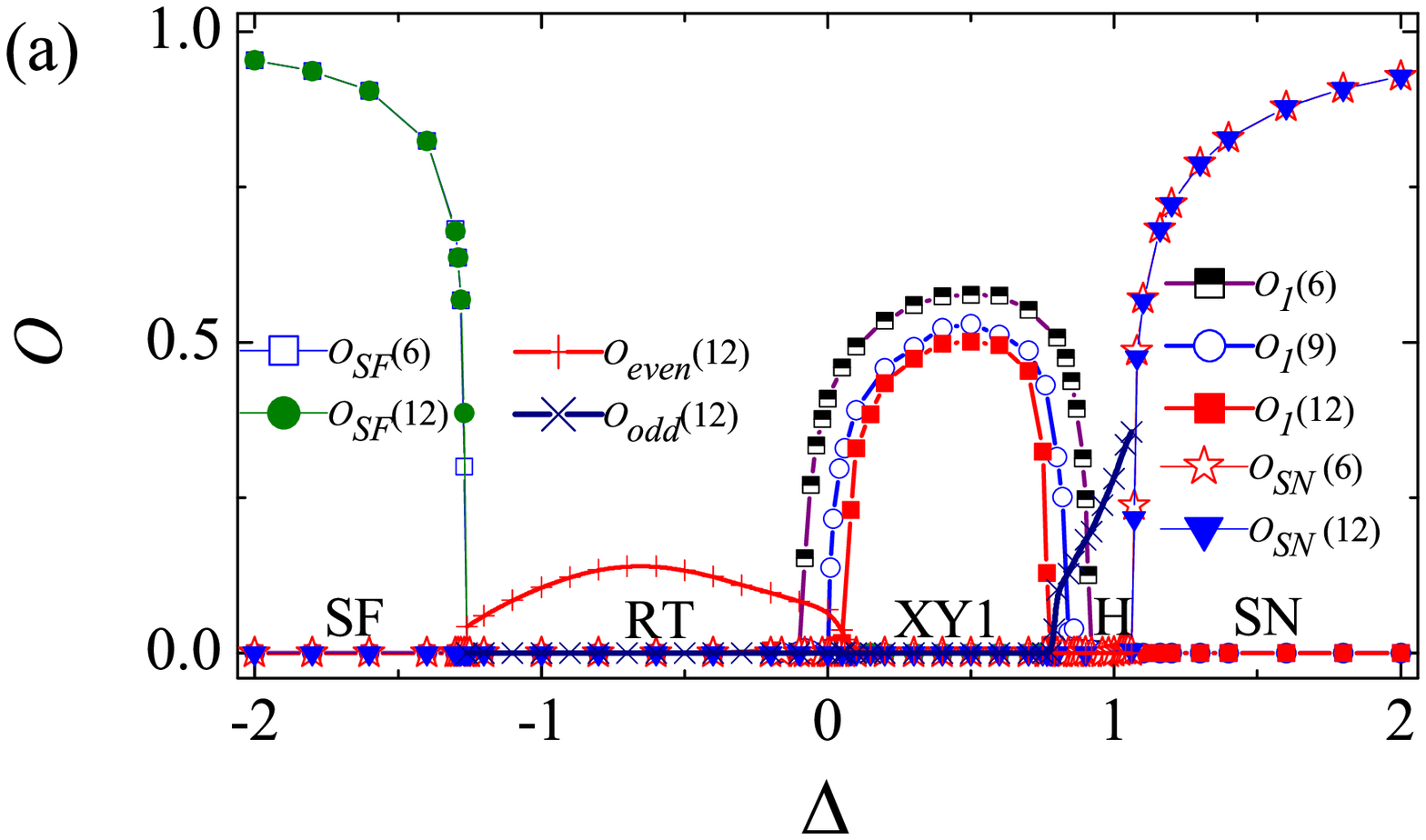}
\includegraphics[width=0.49\textwidth]{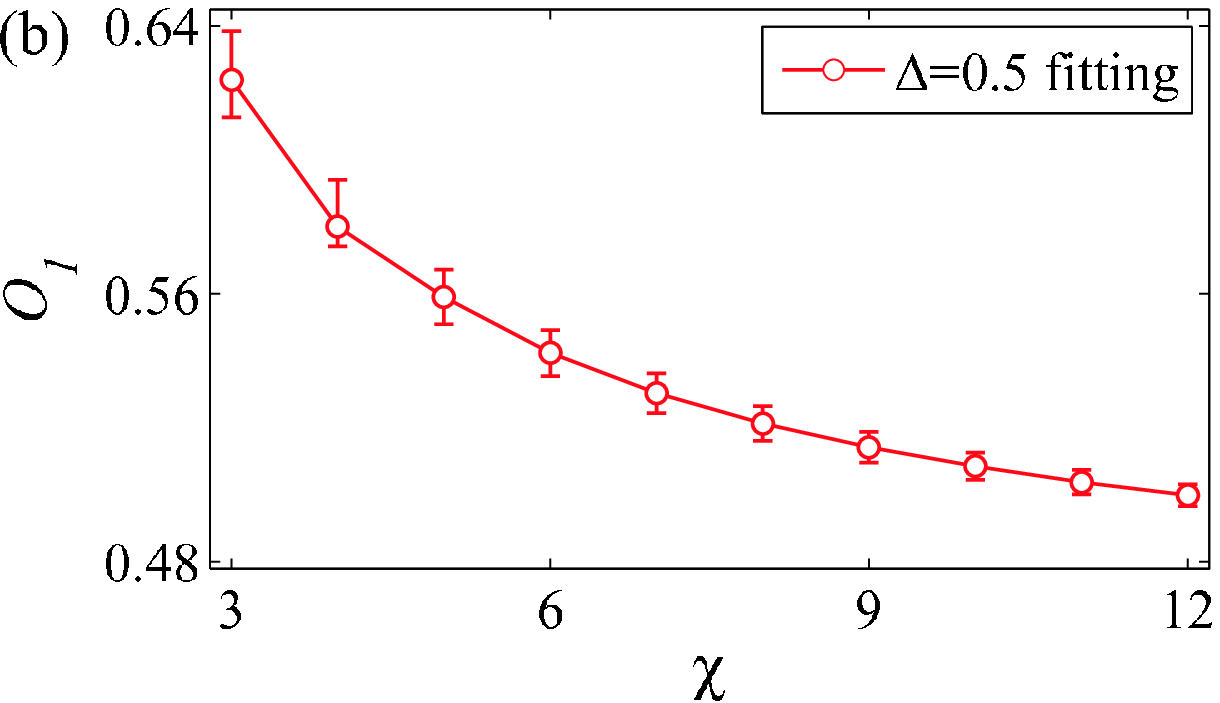}
\includegraphics[width=0.49\textwidth]{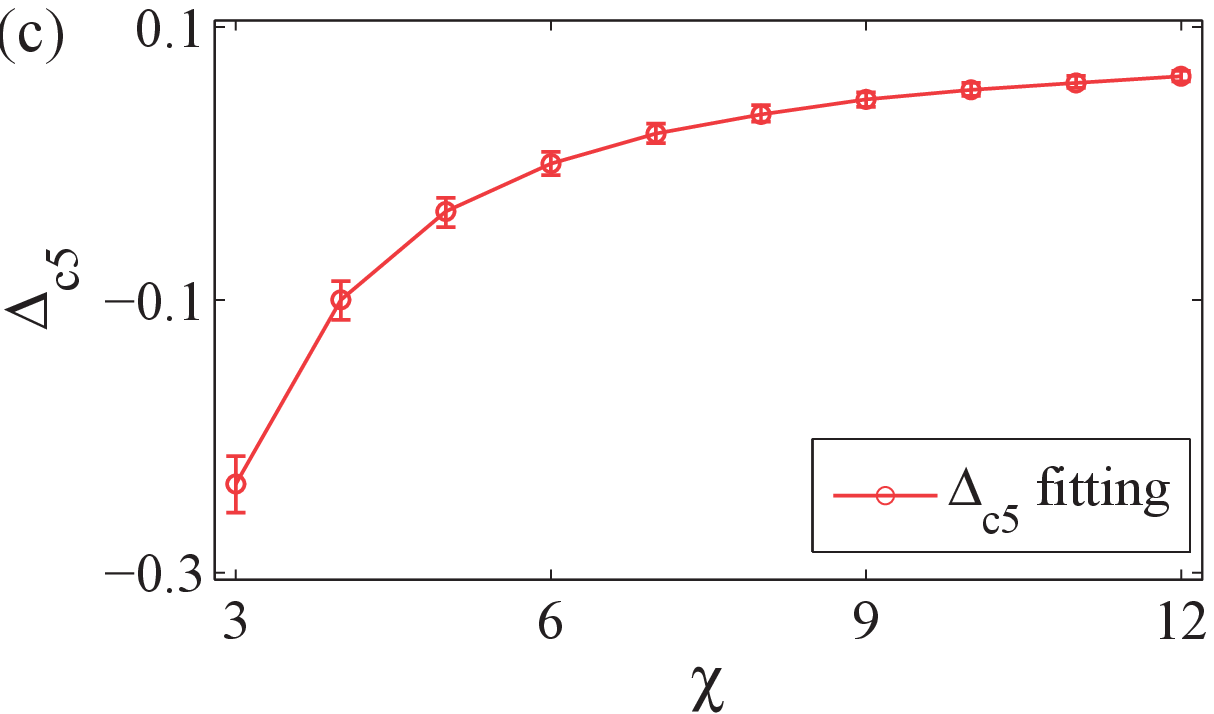}
\includegraphics[width=0.49\textwidth]{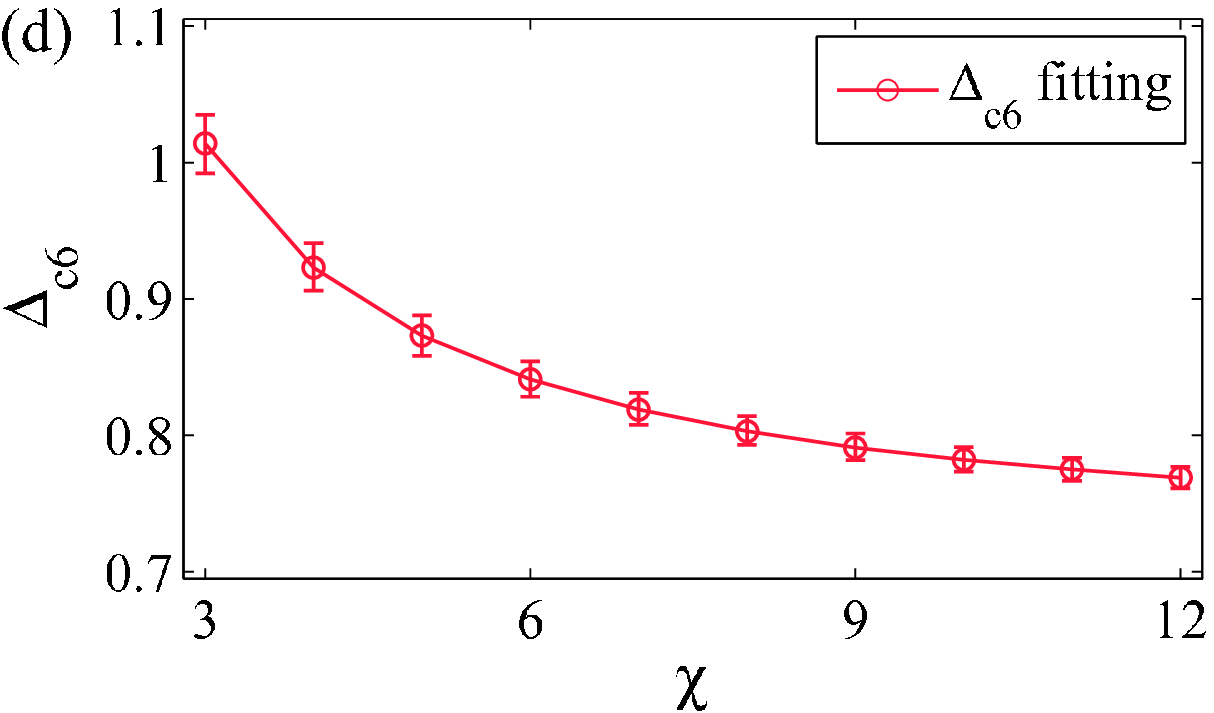}
\end{center}
\caption{
(a) The local order parameters
$O_{\mathrm{SF}}$ in the SF phase and $O_{\mathrm{SN}}$ in the SN phase,
the pseudo-order parameter $O_{1}$ in the $XY1$ phase and the string
order parameters $O_{\mathrm{odd}}$ and $O_{\mathrm{even}}$ in the Haldane and RT ($|1, z\rangle$)
phases as a function of the anisotropy $\Delta$, with $J = -1$.
Two phase transition points are identified at $\Delta_{c4}=-1.26$ and $\Delta_{c7}=1.06$ for $\chi=6$, with
no significant shift observed for increasing $\chi$.
The other two phase transition estimates, $\Delta_{c5}$ and $\Delta_{c6}$, shift with increasing $\chi$.
(b) The pseudo-order parameter estimates $O_{1}(\chi)$ for $\Delta=0.5$ in the $XY1$ phase (see text).
(c) The ``pseudo-critical" point estimates $\Delta_{c5}(\chi)$ obtained from the pseudo-order parameter $O_{1}(\chi)$ and its fitting function (see text).
(d) The ``pseudo-critical" point estimates $\Delta_{c6}(\chi)$ obtained from the pseudo-order parameter $O_{1}(\chi)$ and its fitting function (see text).
}\label{OR2}
\end{figure}

\begin{figure}
\begin{center}
\includegraphics[width=0.49\textwidth]{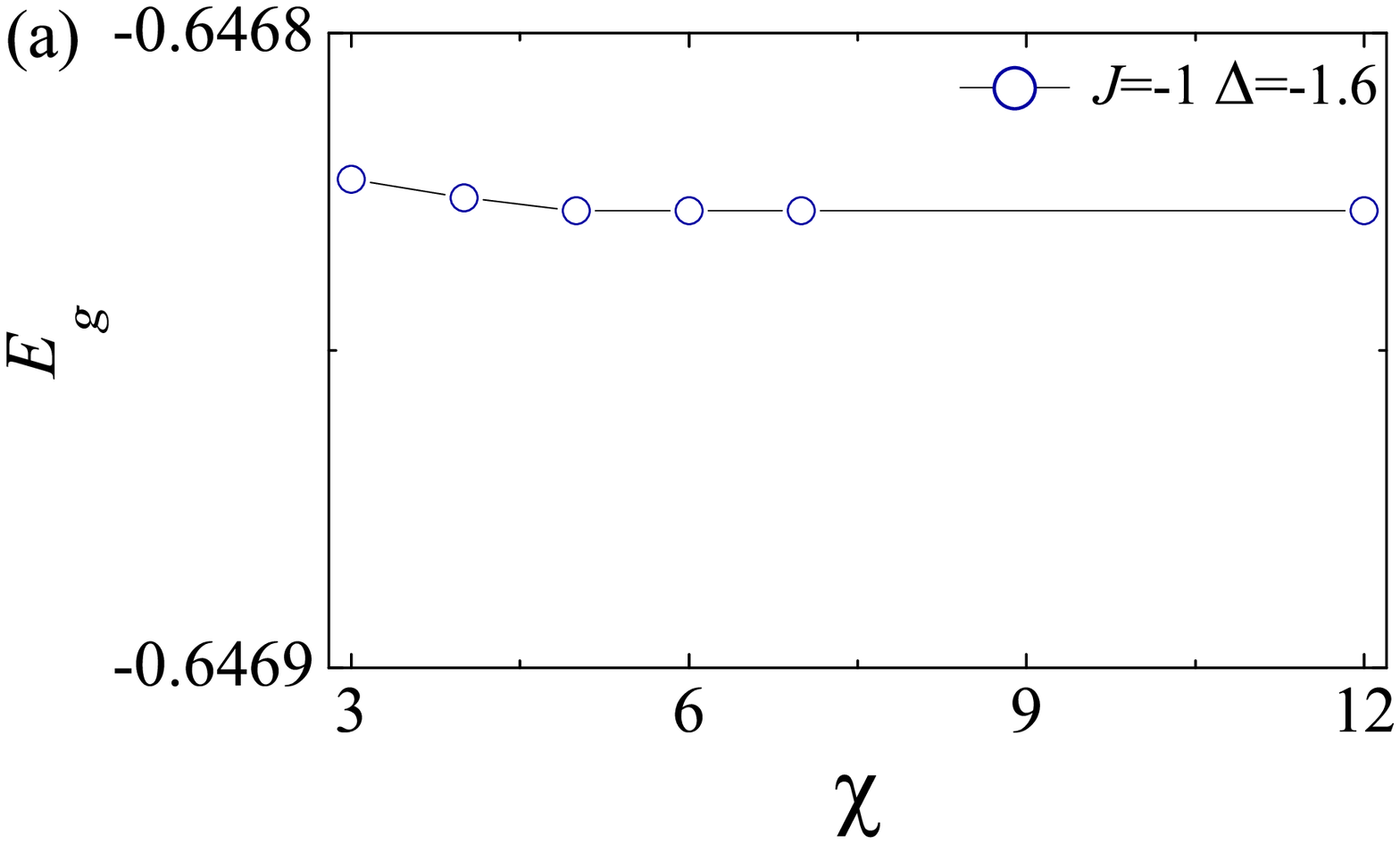}
\includegraphics[width=0.49\textwidth]{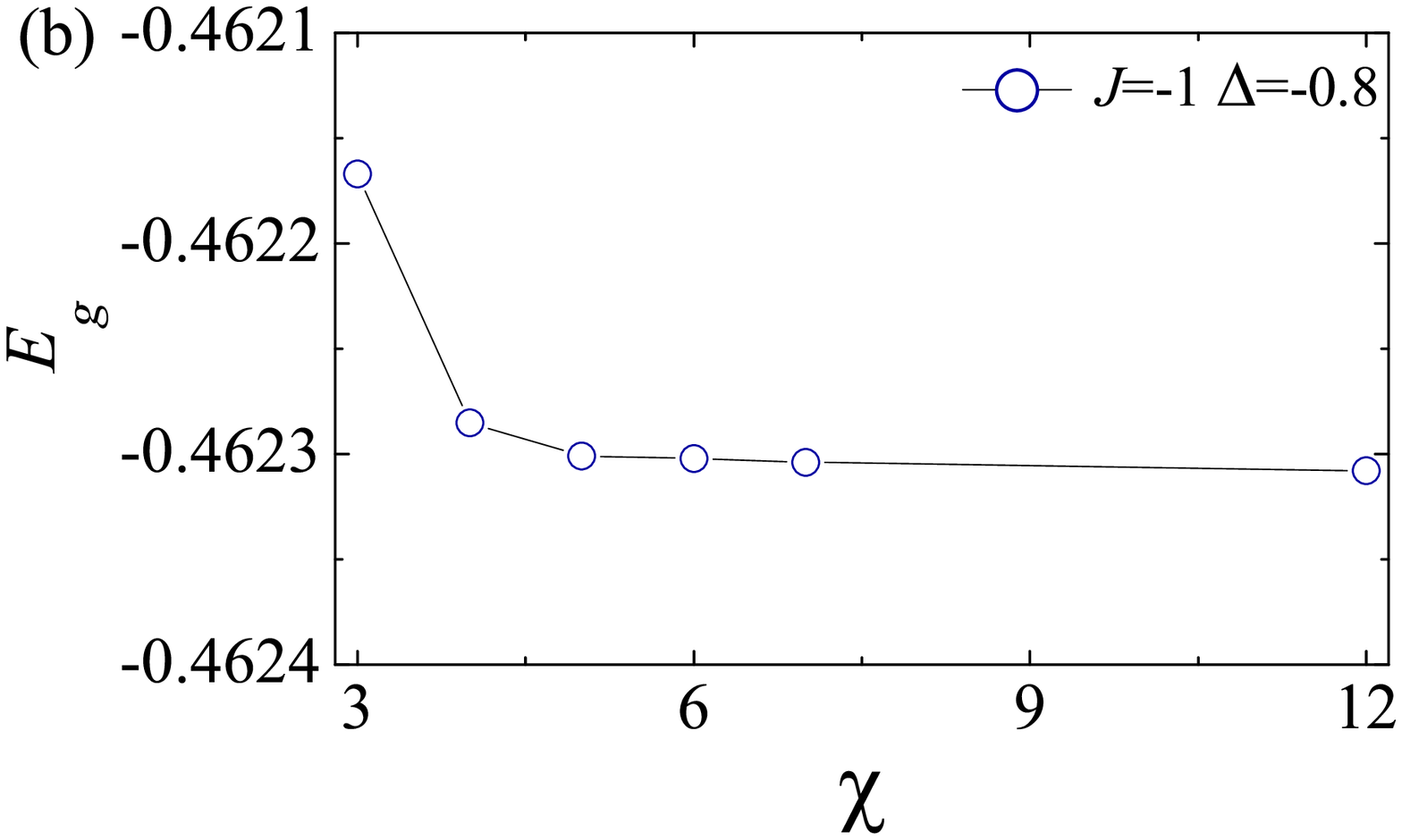}
\includegraphics[width=0.49\textwidth]{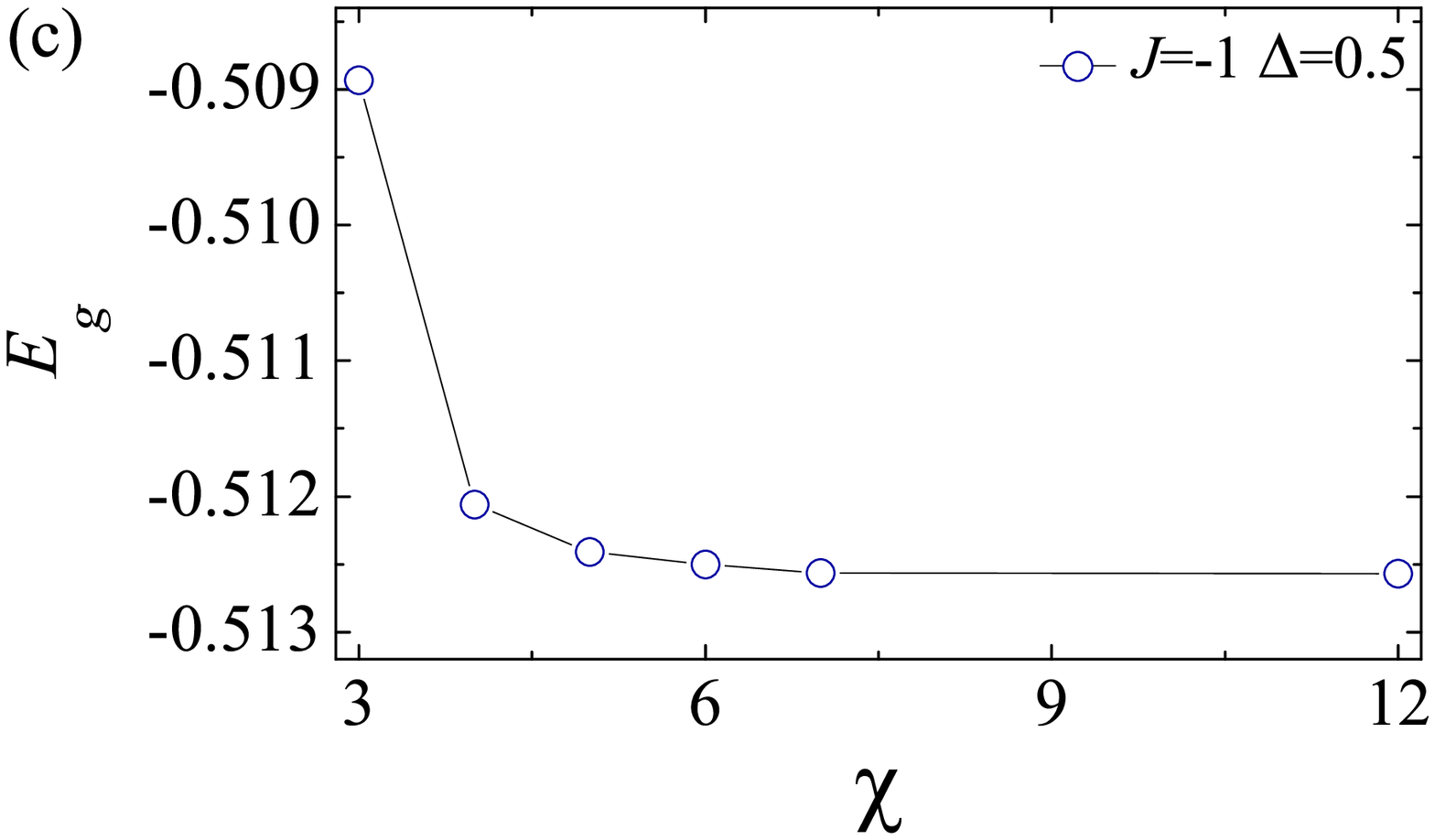}
\includegraphics[width=0.49\textwidth]{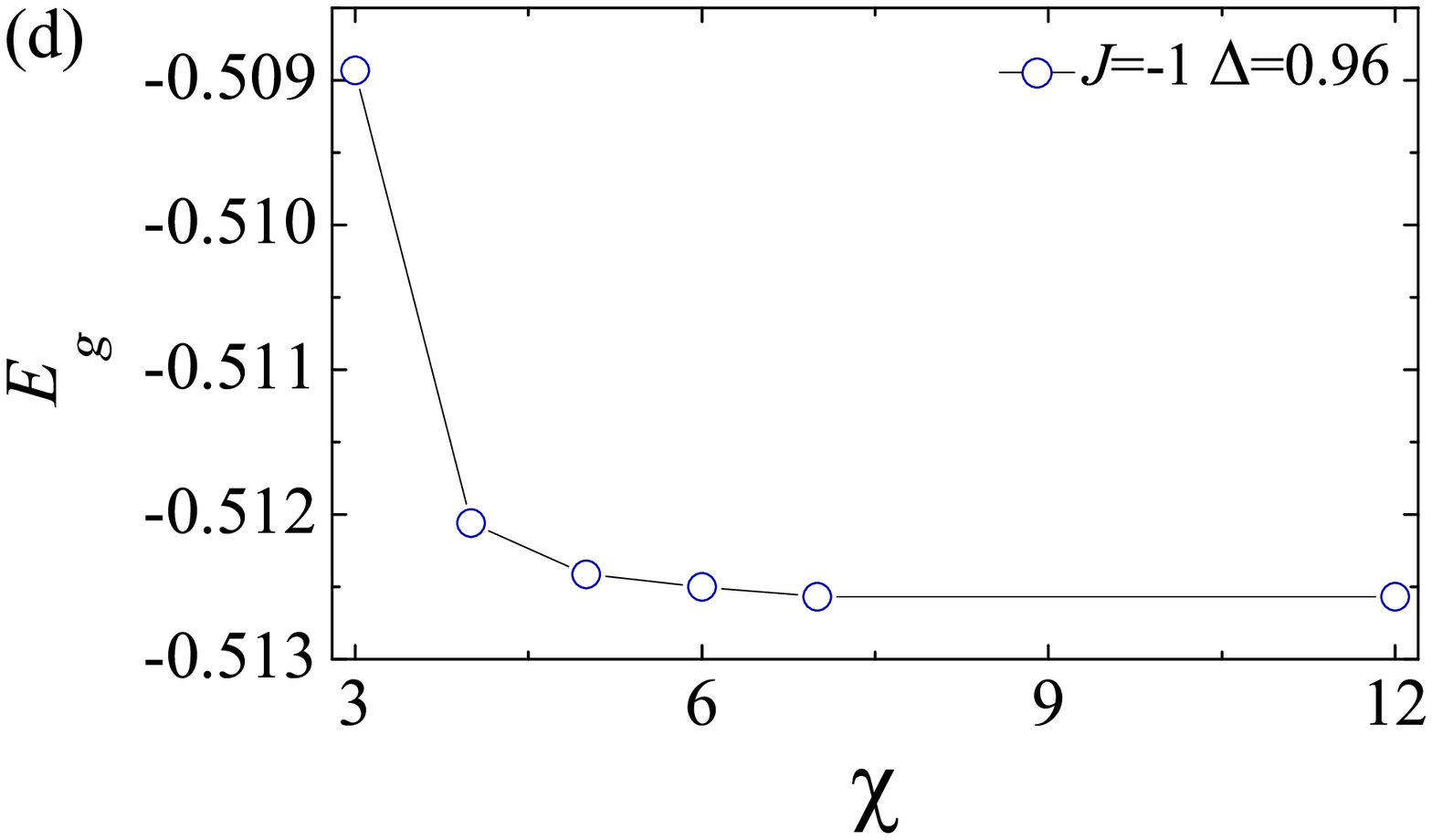}
\includegraphics[width=0.49\textwidth]{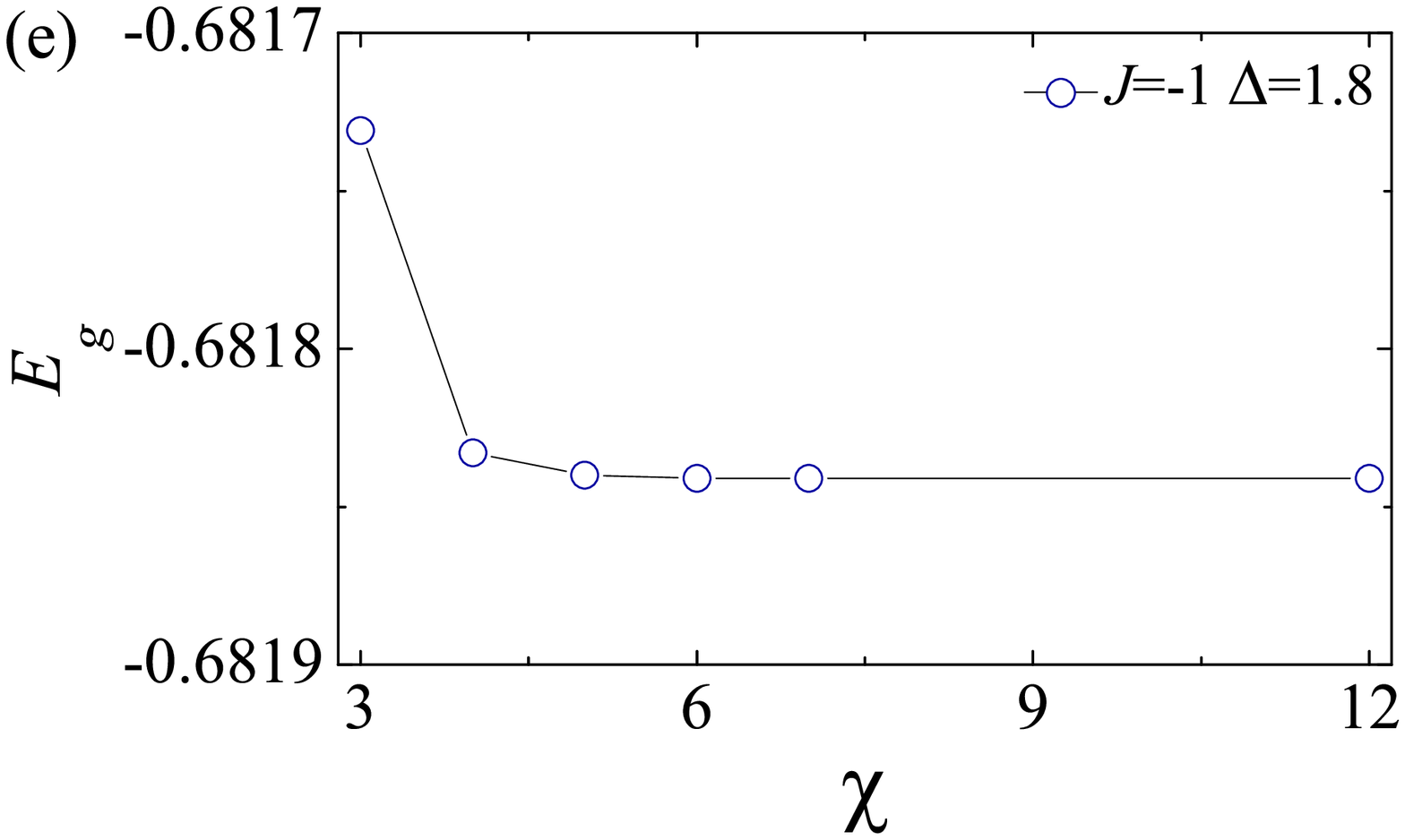}
\end{center}
\caption{Convergence of the groundstate energy $E_g$ with increasing bond dimension $\chi=3,4,5,6,7,12$ in different phases
with rung coupling $J=-1$ and anisotropy parameter values
(a) $\Delta=-1.6$ (SF phase)
(b) $\Delta=-0.8$ (RT phase)
(c) $\Delta=0.5$ ($XY1$ phase)
(d) $\Delta=0.96$ (Haldane phase)
(e) $\Delta=1.8$ (SN phase).
 }\label{Eg2}
\end{figure}

\begin{figure}
\begin{center}
\includegraphics[width=0.6\textwidth]{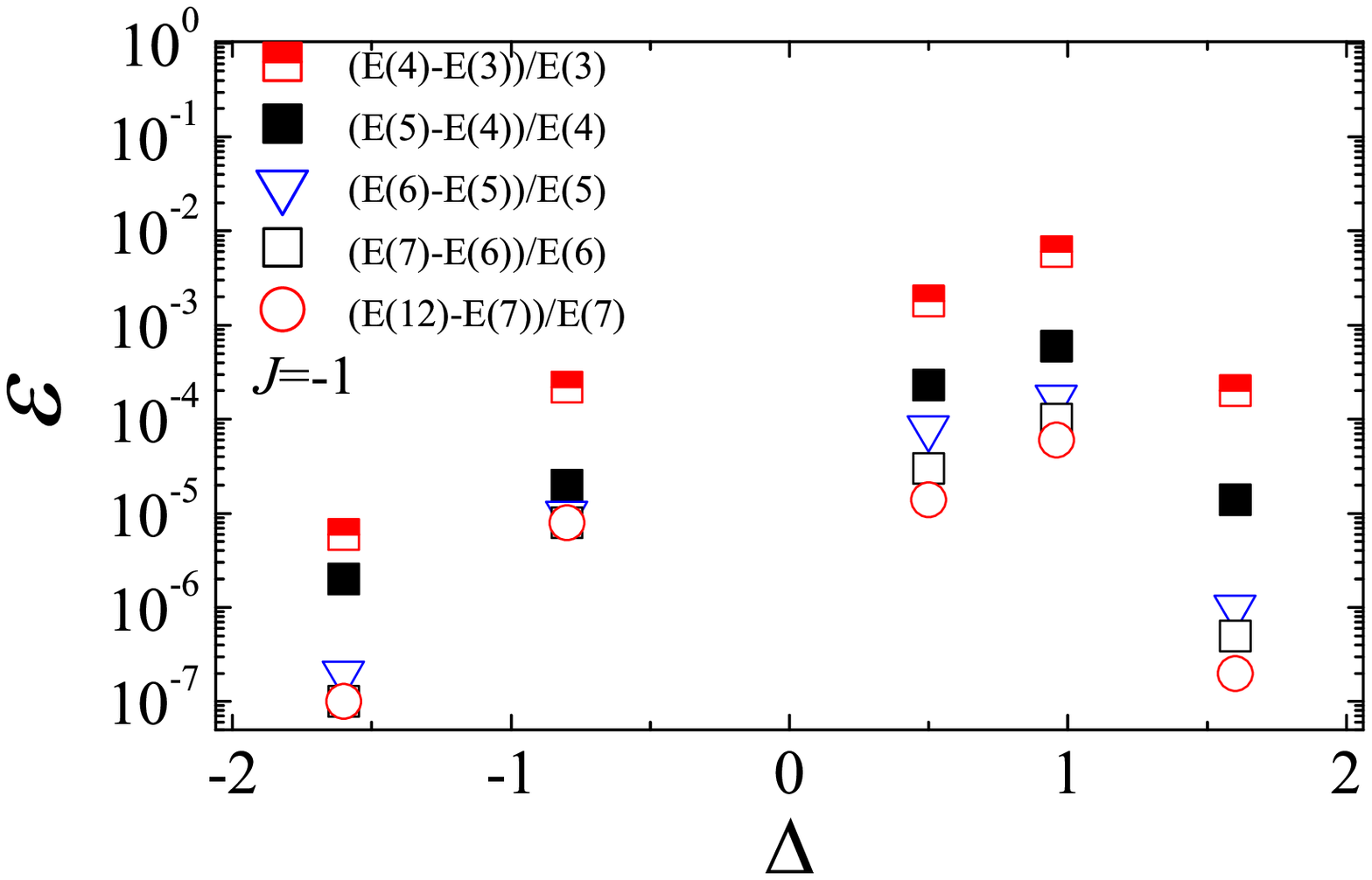}
\end{center}
\caption{Relative error $\varepsilon$ in the estimates of the groundstate energy as a function of $\Delta$ at $J=-1$ for increasing bond dimension $\chi$.
The relative errors are defined by $\varepsilon = (E(\chi_{n+1})-E(\chi_n))/E(\chi_n)$ where $E(\chi_n)$
is the groundstate energy estimate for given bond dimension $\chi_n$.
 }\label{EEE2}
\end{figure}

\begin{figure}
\begin{center}
\includegraphics[width=0.49\textwidth]{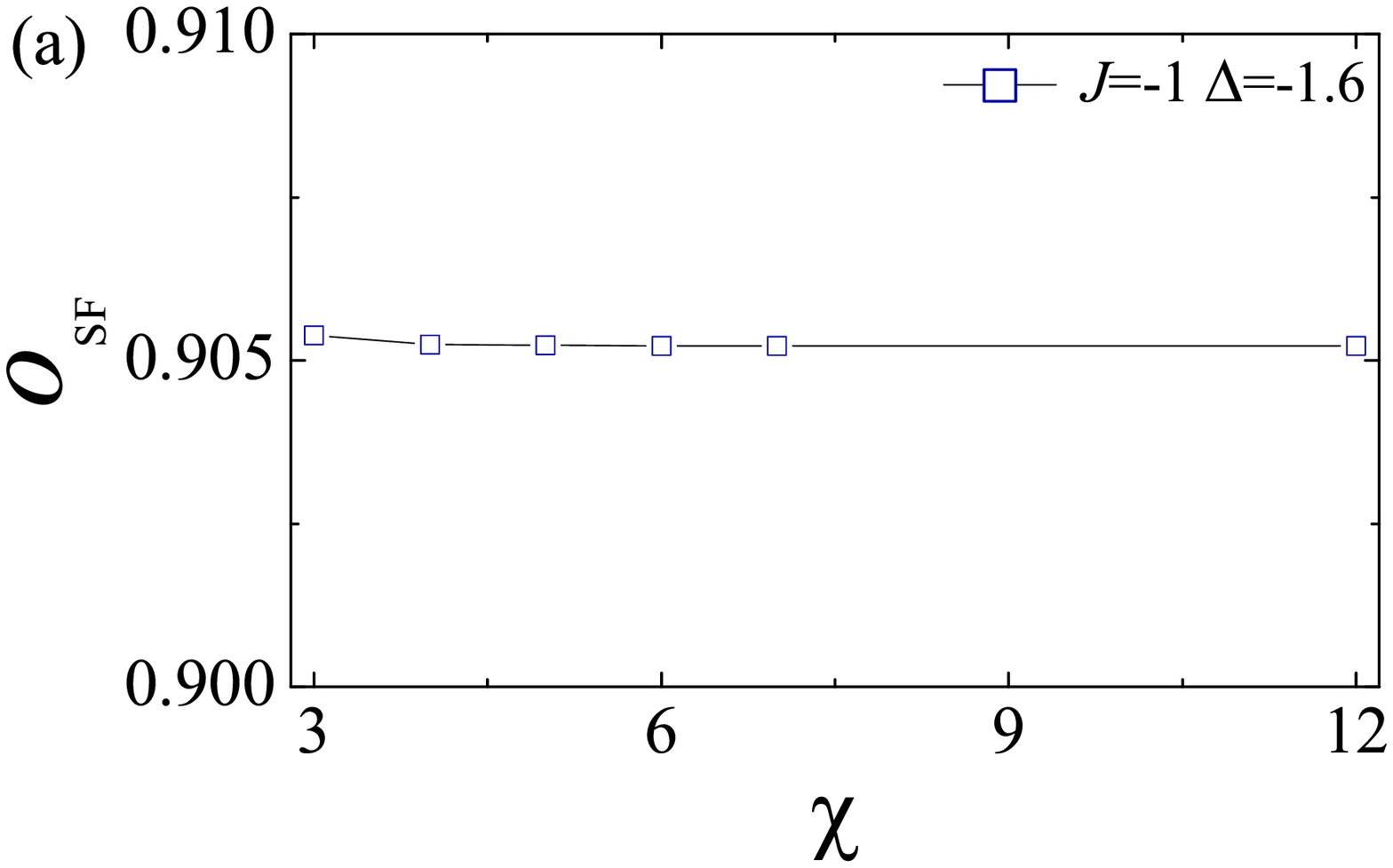}
\includegraphics[width=0.49\textwidth]{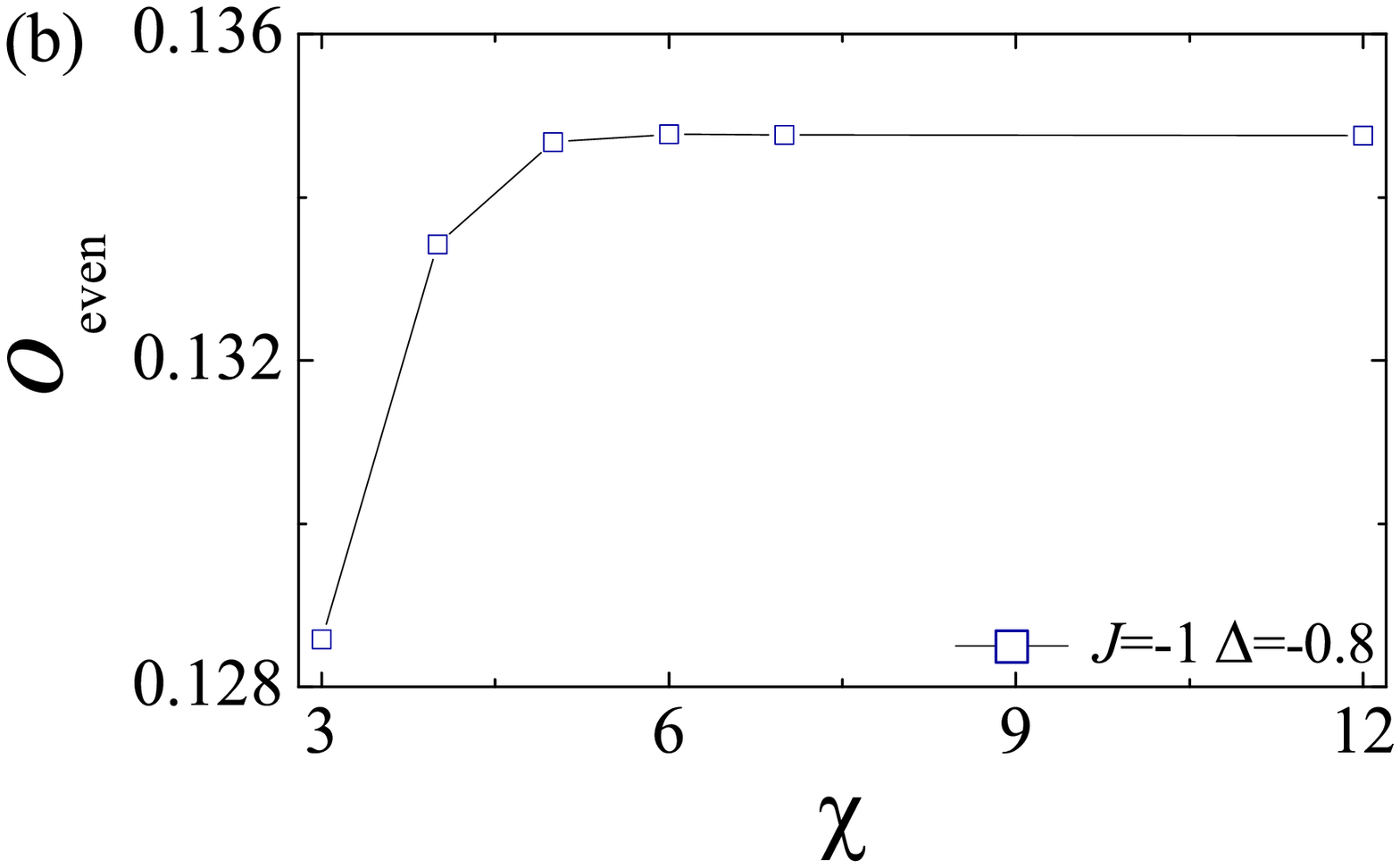}
\includegraphics[width=0.49\textwidth]{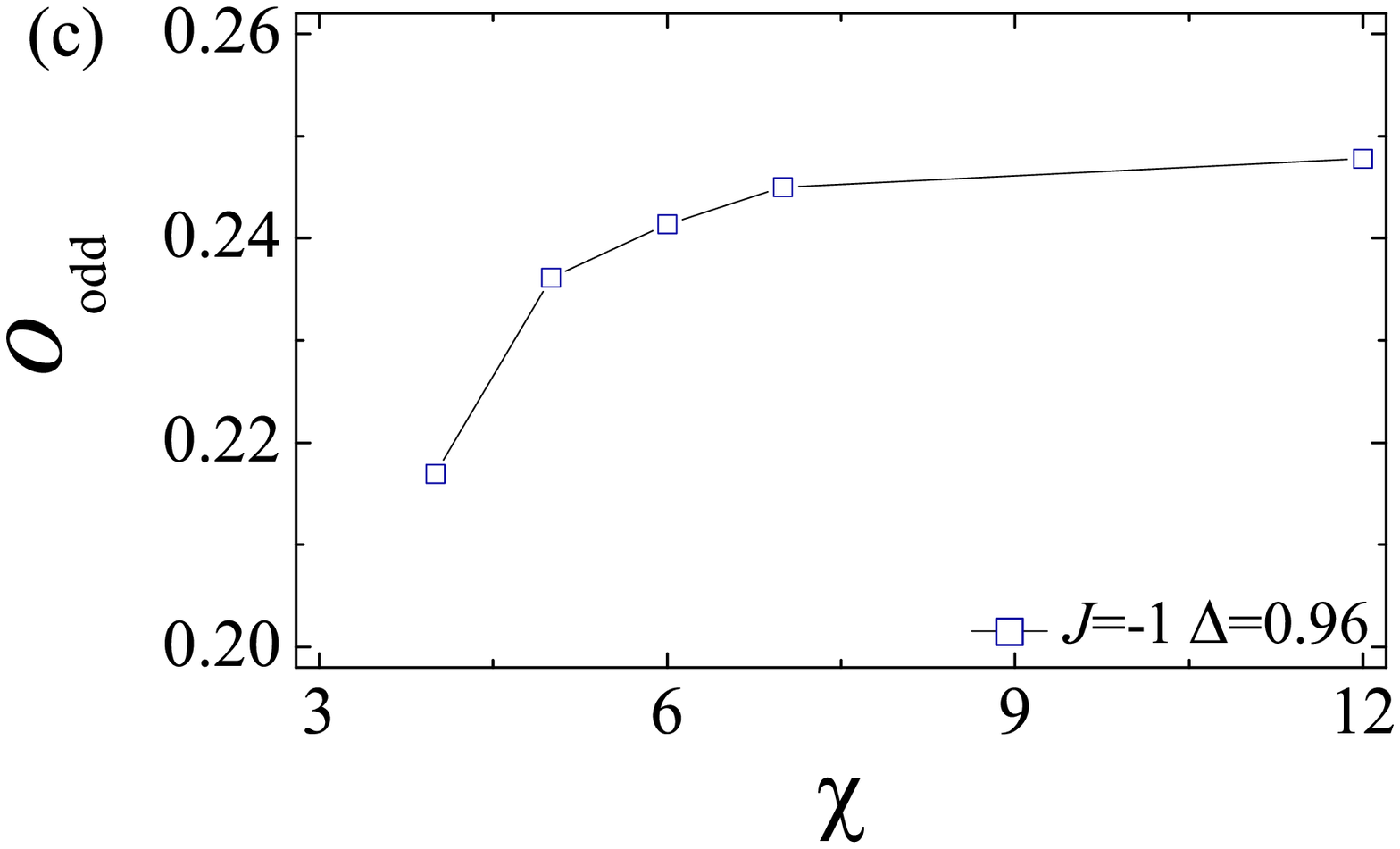}
\includegraphics[width=0.49\textwidth]{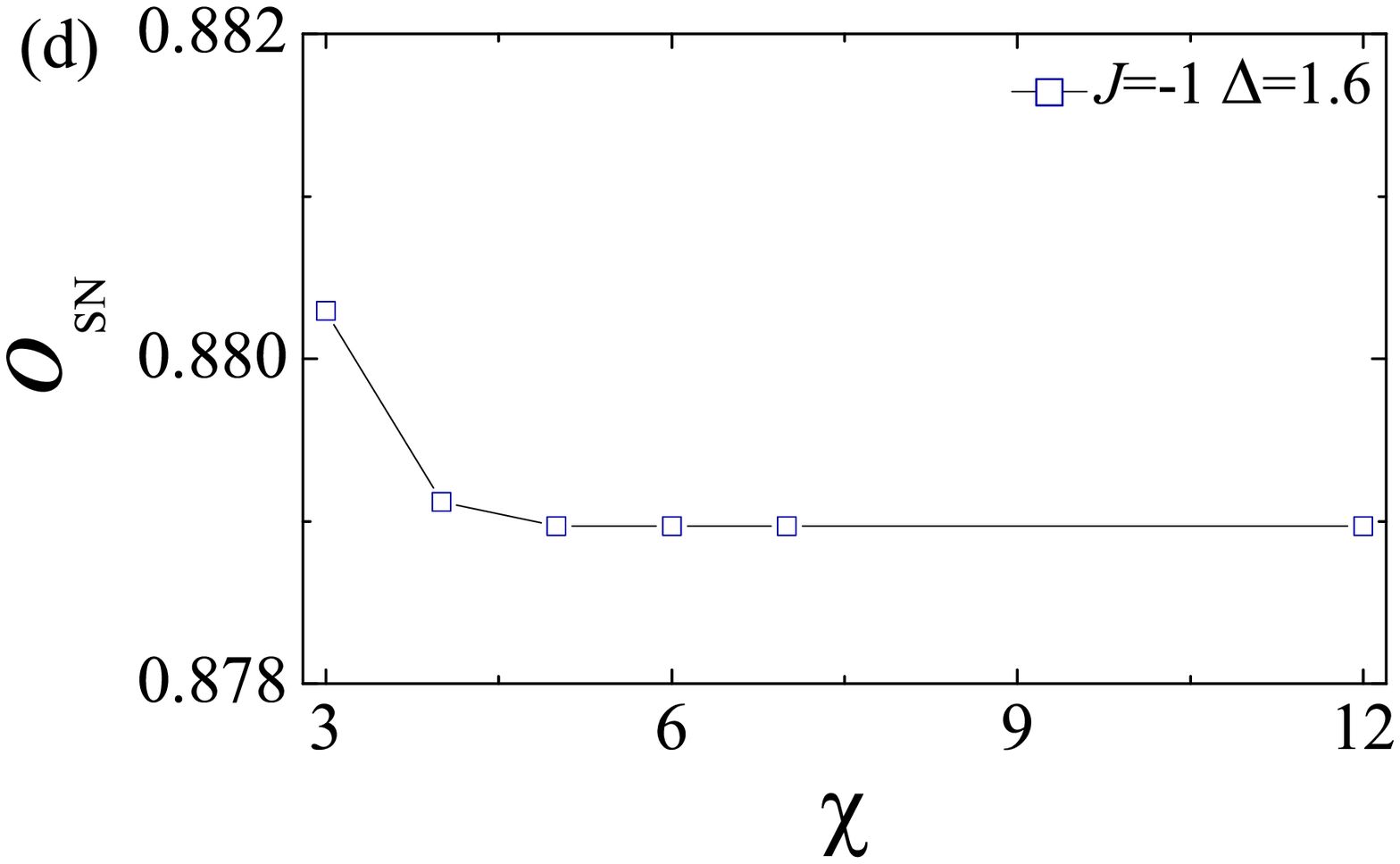}
\end{center}
\caption{Convergence of order parameters with increasing bond dimension $\chi=3,4,5,6,7,12$ in different phases for $J=-1$.
(a) local order parameter $O_{\mathrm{SF}}$ for $\Delta=-1.6$ (SF phase).
(b) string order parameter $O_{\mathrm{even}}$ for $\Delta=0.8$ (RT phase).
(c) string order parameter $O_{\mathrm{odd}}$ for $\Delta=0.96$ (H phase).
(d) local order parameter $O_{\mathrm{SN}}$ for $\Delta=1.6$ (SN phase).
}\label{O2}
\end{figure}

In similar fashion, we determine the phase boundaries between the SF, RT ($|1,z\rangle$), $XY1$, H and SN
phases.
The groundstate fidelity per lattice site $d(\Delta_1,\Delta_2)$ is shown
as a function of  $\Delta_1$ and $\Delta_2$ in Figure~\ref{DDD2} with bond dimension $\chi=6$.
In this region four pinch points, $(\Delta_{c4},\Delta_{c4})$, $(\Delta_{c5},\Delta_{c5})$, $(\Delta_{c6},\Delta_{c6})$ and
$(\Delta_{c7},\Delta_{c7})$, are identified on the fidelity surfaces, corresponding to
the continuous phase transition points $\Delta_{c4}= -1.26$,
$\Delta_{c5}= 0.00$, $\Delta_{c6}= 0.84$ and $\Delta_{c7}= 1.06$.
These points separate five different phases, which we now proceed to characterize in terms of the order parameters.

In the SF phase, the one-site reduced density matrix
yields the local order parameter
\begin{equation}\label {fitlader4}
O_{\mathrm{SF}}=\frac{1}{2} |\langle (S^z_{1, i}- S^z_{2, i})+(S^z_{1,i+1}-S^z_{2, i+1}) \rangle |.
\end{equation}
Our results show that the ladder system exhibits a long-range order in the SF phase for $\Delta < -1.26$
(see Figure~\ref{OR2}(a)).

In the SN phase, we are similarly able to consider the local order parameter
\begin{equation}\label {fitlader5}
O_{\mathrm{SN}}=\frac{1}{2} |\langle (S^z_{1, i}+ S^z_{2, i})-(S^z_{1,i+1}+S^z_{2, i+1}) \rangle|.
\end{equation}
In this phase the ladder system exhibits a long-range order for $\Delta > 1.06$ (see Figure~\ref{OR2}(a)).

In the H phase, the even string order parameter $O_{\mathrm{even}}$ vanishes, but the odd string order parameter
is nonzero.
Conversely, in the RT ($|1, z\rangle$) phase,  $O_{\mathrm{odd}}=0$ and the even string order parameter is nonzero
(see Figure~\ref{OR2}(a)).

In the $XY1$ phase, the pseudo-order parameter $O_{1}$ as a function of
the anisotropy $\Delta$ for different bond dimension $\chi$ is shown in
Figure~\ref{OR2}(a).
The pseudo-order parameter is plotted as a function of $\chi$ for $\Delta=0.5$ in Figure~\ref{OR2}(b).
Here we have performed an extrapolation with respect to $\chi$, with the fitting
function $O_1(\chi)=a_3 \, \chi^{-b_3}(1+c_3 \, \chi^{-1})$, where $a_3=0.467(7)$, $b_3=0.006(4)$ and $c_3=1.04(5)$.
This is again consistent with previous studies of pseudo-order parameters.
The estimates of the ``pseudo-critical" points $\Delta_{c5}(\chi)$ and $\Delta_{c6}(\chi)$ on either side of the $XY1$ phase
are shown in Figure~\ref{OR2}(c) and Figure~\ref{OR2}(d).
Here we employ the extrapolation functions $\Delta_{c5}(\chi)=a_4+b_4 \, \chi^{-c_4}$
and $\Delta_{c6}(\chi)=a_5+b_5 \, \chi^{-c_5}$.
Numerical fitting yields $a_4=0.088(2)$, $b_4=-2.54(7)$ and $c_4=1.88(2)$ for the phase
transition between the RT ($|1, z\rangle$) and $XY1$ phases with the values
$a_5=0.718(2)$, $b_5=1.18(4)$ and $c_5=1.26(3)$ for the phase transition between
the $XY1$ and H phases.
In the limit $\chi\rightarrow\infty$ this yields the two critical point estimates $\Delta_{c5}$ and $\Delta_{c6}$.

It follows that we have been able to compute the order parameters $O_{\mathrm{SF}}$ and
$O_{\mathrm{SN}}$ in the SF and SN phases, the pseudo-order
parameter $O_{1}$ in the $XY1$ phase, the string order parameter
$O_{\mathrm{even}}$ in the RT ($|1, z\rangle$) phase, and the string order parameter
$O_{\mathrm{odd}}$ in the H phase.
The results are shown in Figure~\ref{OR2}.
With the anisotropy $\Delta$ as control parameter, the ladder system undergoes four continuous phase
transitions at the values $\Delta_{c4}$, $\Delta_{c5}$, $\Delta_{c6}$ and $\Delta_{c7}$.

In the limit $J\rightarrow -\infty$ the $XY1$ phase and the H phase vanish.
We note that the phase boundary between the SF phase and the RT ($|1, z\rangle$) phase is located roughly near the
line $\Delta \approx J$ in the limits $\Delta\rightarrow -\infty$ and $J\rightarrow -\infty$,
while the phase boundary between the RT ($|1,z\rangle$) and SN phases is located roughly
near the line $\Delta \sim 1$ as $J\rightarrow -\infty$.
As a result there are three phases for the ladder system in the limit $J\rightarrow -\infty$,
i.e., the SF, RT ($|1,z\rangle$) and SN phases.

The accuracy of the results is similar to that for $J >0$, with estimates of the groundstate energy per site with increasing bond dimension
shown in Figure~\ref{Eg2}.
The estimates also converge rapidly, as quantified in Figure~\ref{EEE2}.
The corresponding convergence of the order parameter estimates is shown in Figure~\ref{O2}.

\begin{figure}
\begin{center}
\includegraphics[width=0.7\textwidth]{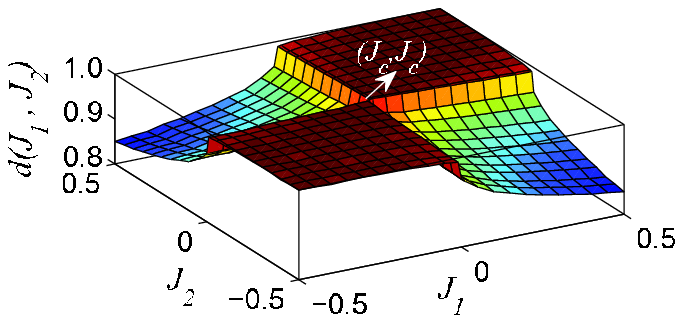}
\end{center}
\caption{The groundstate fidelity surface $d(J_{1},J_{2})$ for
the spin-$\case12$ $XXZ$ two-leg ladder model (\ref{ham1}) for anisotropy parameter $\Delta=0$
and varying rung coupling with bond dimension $\chi=6$.
The pinch point occurring at $(J_{c}= 0.00, J_{c}= 0.00)$
indicates a continuous phase transition point.}
\label{DDD3}
\end{figure}

\begin{figure}
\begin{center}
\includegraphics[width=0.6\textwidth]{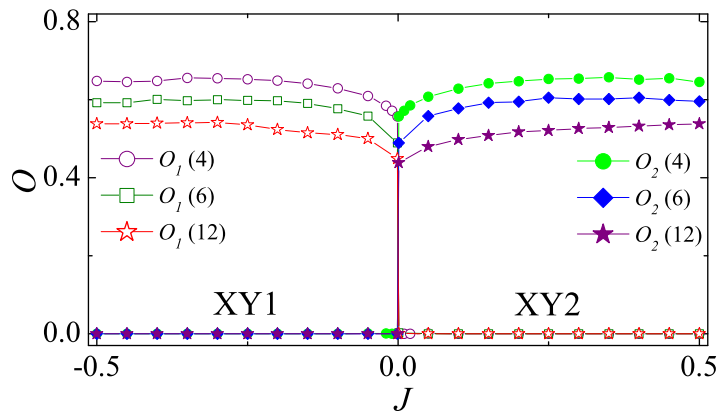}
\end{center}
\caption{The pseudo-order parameter $O_{1}(\chi)$ in the $XY1$ phase and the pseudo-order
parameter $O_{2}(\chi)$ in the $XY2$ phase as a function of the rung coupling $J$.
The bond dimension values are $\chi=4, 6, 12$.}
\label{OR3}
\end{figure}

\subsection{$\Delta=0, -0.5 \le J \le 0.5$}

We now turn to look more closely at the $XY$ phase, which as we have seen, splits into the two different
$XY1$ and $XY2$ phases identified by the local pseudo-order parameters $O_1$ and $O_2$ obtained
from the groundstate fidelity per lattice site. As we shall see, for $-1\leq\Delta\leq1$,
the phase boundary between the $XY1$ and $XY2$ phases is the line $J=0$.
Here we fix the anisotropy coupling $\Delta=0$ and take $J$ as control parameter in the range $-0.5 \le J \le 0.5$
to sweep across this boundary.

Figure~\ref{DDD3} shows a plot of the groundstate fidelity per site $d(J_{1},J_{2})$ with bond dimension $\chi=6$.
In this case there is a clear indication of a continuous phase transition point at $J= 0.00$,
as a pinch point occurs at this value with bond dimension $\chi=6$.
Figure~\ref{OR3} shows a plot of the pseudo-order parameter $O_{1}$ in the $XY1$ phase
and the  pseudo-order parameter $O_{2}$ in the $XY2$ phase as a function of $J$.
As already remarked, a feature of the pseudo-order parameters is that they vanish as the bond dimension $\chi$ increases.

\section{Conclusion}

We have systematically investigated the groundstate phase diagram of the infinite fully anisotropic spin-$\case12$ $XXZ$ two-leg
ladder model (\ref{ham1}).
This has been achieved by exploiting a TN algorithm tailored to the geometry of quantum spin ladder systems~\cite{lsh},
allowing efficient evaluation of the groundstate fidelity per lattice site from the TN representation of the
groundstate wave functions for the infinite-size quantum spin ladder.
Illustrative results  have been presented in Sec.~III
along the lines $J=\pm1$ with varying $\Delta$ and the line $\Delta=0$ with varying $J$.
Our computational results using the groundstate fidelity per lattice site are consistent
with results obtained from the construction of local, and where appropriate, nonlocal, order parameters in the different phases.

The full phase diagram, as obtained by this approach, is shown in Figure~\ref{phase}.
This phase diagram, featuring nine distinct quantum phases, points to the rich diversity of physics
in the fully anisotropic spin-$\case12$ $XXZ$ two-leg ladder model.
The precise nature of the phase diagram is a significant extension on the previously
obtained schematic phase diagram for this model, which was mapped out by identifying
ridges and valleys in contour plots of the rescaled block entanglement entropy
corresponding to possible phase boundaries~\cite{XXZ2}.
With that approach, which was aimed at demonstrating the usefulness of sublattice entanglement entropy as a means of exploring
quantum phase transitions across a range of different spin systems, the nature and number of the different phases of the ladder model were not discussed in detail.
The phase diagram of the fully anisotropic spin-$\case12$ $XXZ$ two-leg ladder model
also differs in significant aspects compared to the phase diagram of the spin-$\case12$ $XXZ$ two-leg ladder model with
isotropic rung interactions, the precise nature of which has been the subject of debate~\cite{p1,XXZ1}.
With isotropic rung interactions, the model also exhibits FM,  SF,  RS,  N, SN, H, $XY1$ and $XY2$ phases.
However, for the fully anisotropic model considered here, there is an additional rung-triplet (RT($|1,z\rangle$)) phase, with the
$XY1$ and $XY2$ phases extending over the whole $-1 \le \Delta \le 1$ region in the vicinity of $J \approx 0$.
Moreover, the extent of the Haldane (H) phase is seen to be significantly diminished for fully anisotropic interactions.
In general the difference between both the number of quantum phases and the phase diagrams of the two models highlights the key role played by anisotropies in the
competition between singlet formation and magnetic ordering in quantum spin systems.

\ack

We thank Sam Young Cho, Bing-Quan Hu and Fabian Essler for helpful discussions at various stages of this work.
The work is supported by Chongqing Research Program of Basic Research and Frontier Technology (Grant No. cstc2016jcyjA0480),
National Natural Science Foundation of China (Grant No. 11575037, 11374379, 11174375)
and by the Science and Technology Research Program of Chongqing Municipal Education
Commission (Grant No. KJ1732433).

\appendix

\section{Tensor network representation for the spin-$\case12$ two-leg ladder}

\begin{figure}
\begin{center}
\includegraphics[width=0.9\textwidth]{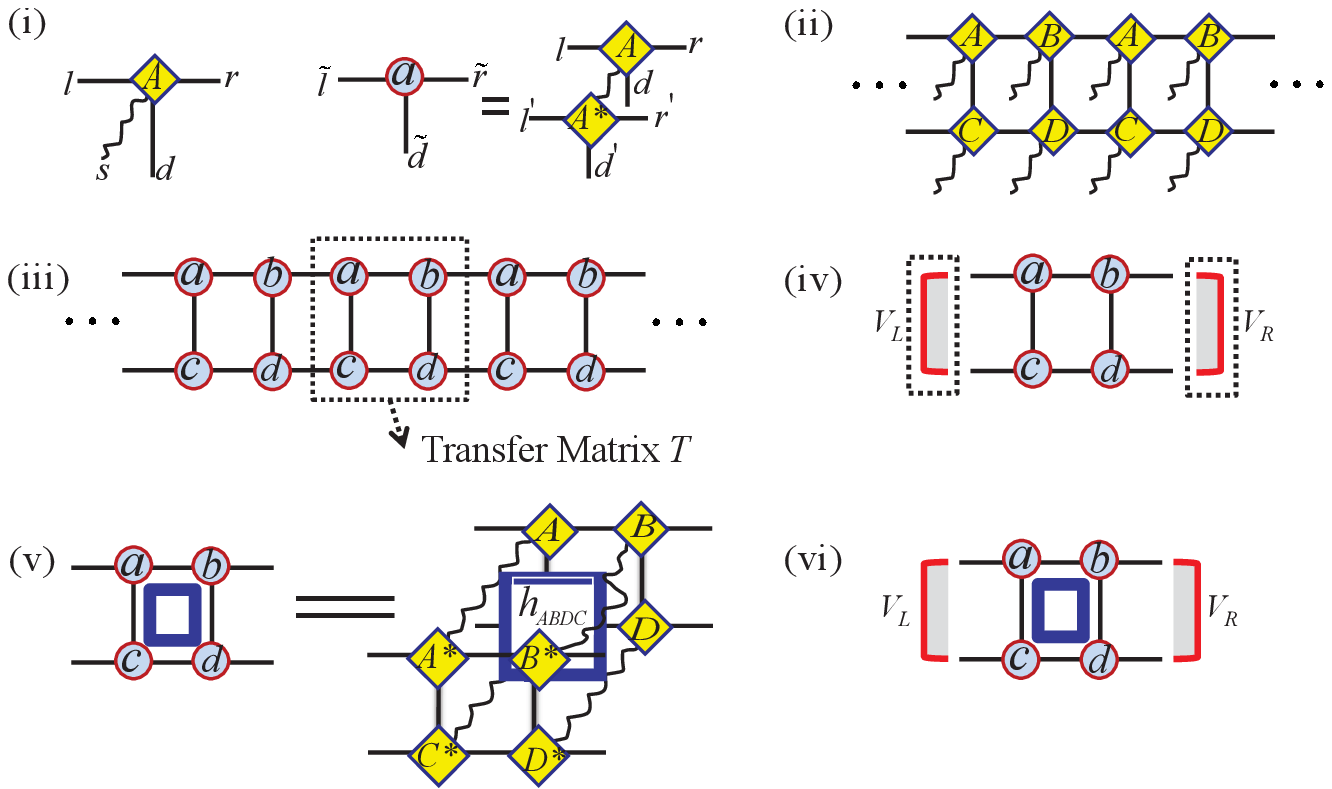}
\end{center}
\caption{
(i) Four-index tensor $A^{s}_{lrd}$ used in the TN representation
for the groundstate wave function for the infinite-size two-leg spin ladder.
Also shown is the double tensor
$a_{\tilde{\ell} \tilde{r}\tilde{d}}$ formed from the
tensor $A_{\ell rd}^{s}$ and its complex conjugate ${(A^*)_{\ell' r' d'}^{s}}$,
with $\tilde{\ell}=(\ell,\ell')$, $\tilde{r}=(r,r')$ and $\tilde{d}=(d,d')$.
(ii) Pictorial representation for a TN state ${| \psi \rangle}$ with leg and rung bonds,
which are used to absorb an operator acting on the $i$th plaquette.
(iii) TN representation for the norm of a groundstate wavefunction in the infinite ladder.
Also shown is the transfer matrix $T$, which is constructed from the four double
tensors $a_{\tilde{\ell} \tilde{r}\tilde{d}}$, $b_{\tilde{\ell}\tilde{r}\tilde{d}}$, $c_{\tilde{\ell} \tilde{r}\tilde{u}}$ and
$d_{\tilde{\ell} \tilde{r}\tilde{u}}$, with $\tilde{\ell}$, $\tilde{r}$ and $\tilde{d}$ defined as above and $\tilde{u}=(u,u')$.
(iv) The dominant left and right eigenvectors $V_{L}$ and $V_{R}$ of the transfer matrix $T$.
(v) A unit cell with hamiltonian density $h_{ABDC}$ acting on the plaquette ABDC.
(iv) The groundstate energy per unit cell is computed from the eigenvectors $V_L$, $V_R$ and the tensors
$A_{\ell rd}^{s}$, $B_{\ell rd}^{s}$, $C_{\ell ru}^{s}$ and $D_{\ell ru}^{s}$.
}
\label{TNtwoleg}
\end{figure}

In this Appendix we summarise the relevant details of the TN representation for spin ladders~\cite{lsh}.
The hamiltonian  of the two-leg ladder model (\ref{ham1}) under consideration is of the general form
$H=\sum_i h_i$, with $h_i$ acting on the $i$th plaquette between sites $i$ and $i+1$ along a leg.
The index $i$ runs over all possible plaquettes with $i \in \{-\infty, \ldots,+\infty \}$.
It is sufficient to assume that the TN representation for the wave function is translational invariant under shifts by two
lattice sites along the legs.
The groundstate wave function for an infinite length ladder can be described by the four
different four-index tensors $A_{\ell rd}^{s}$, $B_{\ell rd}^{s}$, $C_{\ell ru}^{s}$ and $D_{\ell ru}^{s}$.
The tensor $A_{\ell rd}^{s}$ is depicted in Figure~\ref{TNtwoleg}(i),
with $s=1,2$ for spin-$\case12$ and with $\ell$, $r$, $u$, $d=1,\ldots,\chi$, the four inner bond indices, where
$\chi$ is the bond dimension.
A TN representation for the groundstate wave function is shown in Figure~\ref{TNtwoleg}(ii)
with one of the two equivalent choices shown for the plaquette (unit cell).

A gradient-directed random walk method is applied to compute the groundstates.
The double tensors
$a_{\tilde{\ell} \tilde{r}\tilde{d}}$, $b_{\tilde{\ell} \tilde{r}\tilde{d}}$, $c_{\tilde{\ell} \tilde{r}\tilde{u}}$ and
$d_{\tilde{\ell} \tilde{r}\tilde{u}}$ are introduced, with
$\tilde{\ell}=(\ell,\ell')$, $\tilde{r}=(r,r')$, $\tilde{u}=(u,u')$ and $\tilde{d}=(d,d')$.
They are formed from the tensors $A_{\ell rd}^{s}$, $B_{\ell rd}^{s}$, $C_{\ell ru}^{s}$ and $D_{\ell ru}^{s}$,  and their
complex conjugates, as depicted in Figure~\ref{TNtwoleg}(i) for the double tensor $a$.
With these double tensors, the TN for the norm of a wave function is shown in Figure~\ref{TNtwoleg}(iii).
The two different but equivalent choices for the transfer matrix $T$ are
made from the plaquettes $abdc$ and  $bacd$.
In the above, the notation $\tilde{\ell}=(\ell,\ell')$ corresponds to the two independent indices $\ell$ and $\ell'$, with
$\tilde{\ell}=1,2,\ldots,\chi^2$.
Similarly for $\tilde{r}$, $\tilde{u}$ and $\tilde{d}$.

For a randomly chosen initial
state $| \psi_0 \rangle$, the energy per plaquette $e$ is
\begin{equation}
e=\frac{\langle \psi_0 | (h_{ABDC}+h_{BACD}) |\psi_0 \rangle}{ 2 \langle \psi_0 |\psi_0
\rangle}.
\end{equation}
The groundstate energy per plaquette also admits a TN
representation as shown in Figure~\ref{TNtwoleg}(v) and (vi) for plaquette ABDC, which absorbs the operator $h$ acting on a plaquette for
an infinite-size spin ladder.
For each choice of plaquette, the dominant left and right eigenvectors of the transfer matrix $T$
constitute the environment tensors, visualised in Figure~\ref{TNtwoleg}(iv) and (vi) for plaquette ABDC.
In this way the energy per plaquette $e_{ABDC}$ is obtained. The same procedure may
be used to compute the energy per plaquette  $e_{BACD}$  for an operator acting
on the $i$th plaquette BACD.

\begin{figure}
\begin{center}
\includegraphics[width=0.9\textwidth]{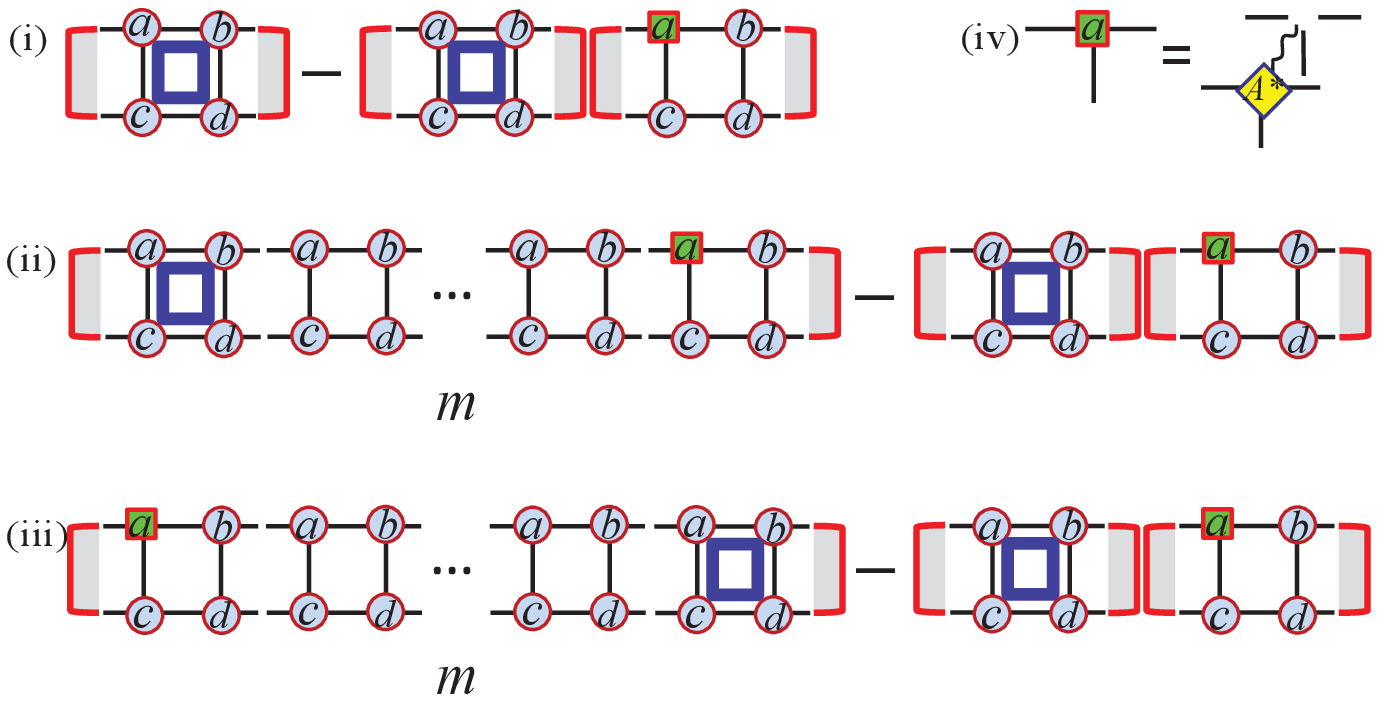}
\end{center}
\caption{The contribution to the energy gradient (\ref{grad}) for the infinite-size spin ladder consists of three
parts: (i), (ii) and (iii) depending on the location of the hole cell
with tensor $A_{\ell rd}^{s}$ removed. In the latter
two cases, there are $m$ cells between the hole and hamiltonian cell, where $m \in (0,1,2,3,\ldots)$.
The hole cell is visualised in (iv), with the tensor $A_{\ell rd}^{s}$ removed. Note
that only the contribution from the plaquette $abdc$, as labeled
here, is shown. There is also a similar contribution from the plaquette $bacd$. }\label{TNupdate}
\end{figure}

To update the TN representation, the energy gradient is computed with
respect to the tensor $A_{\ell rd}^{s}$, with
\begin{equation}
\frac{\partial {e}}{\partial A_{\ell rd}^{s}}=\frac{{\partial
\langle \psi_0 | H |\psi_0 \rangle}/{\partial A_{\ell
rd}^{s}}}{\langle \psi_0 |\psi_0 \rangle}- {E_g} \,
\frac{{\partial \langle \psi_0 |\psi_0 \rangle}/{\partial A_{\ell
rd}^{s}}}{\langle \psi_0 |\psi_0 \rangle}.
\label{grad}
\end{equation}
The contributions to the energy gradient come from three parts (for a given choice of
plaquette on which the hamiltonian acts), as shown in Figure~\ref{TNupdate}.
To obtain the energy gradient it is necessary to sum the contributions from both plaquettes $abdc$ and $bacd$.
The tensor $A_{\ell rd}^{s}$ is then updated according to
\begin{equation}
A_{\ell rd}^{s}=A_{\ell rd}^{s} - \delta \; \frac{\partial
{E_g}/\partial A_{\ell rd}^{s}}{\max_{(\ell, r, d)}\mid\partial
{E_g}/\partial A_{\ell rd}^{s}\mid},
\end{equation}
where $\delta$ denotes the update step size.
In this procedure the tensors $A_{\ell rd}^{s}$, $B_{\ell rd}^{s}$, $C_{\ell ru}^{s}$ and $D_{\ell ru}^{s}$
are updated simultaneously.
Repeating this procedure until the groundstate energy per plaquette converges
leads to the groundstate wave function as approximated in the TN representation.

\end{document}